\documentclass[]{aa}
\newcommand{\vsini}{\ensuremath{{\varv} \sin i}}
\newcommand{\kms}{\ensuremath{\text{km}\,\text{s}^{-1}}} 
\newcommand{\zav}[1]{\left(#1\right)}

\newcommand{\hd}{HD\,60431}
\newcommand{\Teff}{\ensuremath{T_{\mathrm{eff}}}}
\newcommand{\lam}{$\lambda$}
\newcommand{\Ms}{\ensuremath{\mathrm{M}_{\odot}}}
\newcommand{\Rs}{\ensuremath{\mathrm{R}_{\odot}}}
\newcommand{\T}{\mathit{\Theta}}

\newcommand{\Rp}{\ensuremath{R_{\mathrm{p}}}}
\newcommand{\Req}{\ensuremath{R_{\mathrm{eq}}}}
\newcommand{\Om}{\ensuremath{\mathit{\Omega}}}
\newcommand{\bd}{\ensuremath{B_\mathrm{d}}}

\usepackage{graphicx}
\usepackage{txfonts}
\usepackage[]{hyperref}
\begin{document}

   \title{HD 60431, the CP2 star with the shortest rotational period}
   \subtitle{Physical parameters and period analysis}

   \author{Z.~Mikul\'a\v{s}ek\inst{1}
          \and
          E.~Semenko\inst{2}
          \and
          E.~Paunzen\inst{1}
          \and
          S.~H\"ummerich\inst{3,4}
          \and
          P.~L.~North\inst{5}
          \and
          K.~Bernhard\inst{3,4}
          \and
          J.~Krti\v{c}ka\inst{1}
         \and
         J.~Jan\'{i}k\inst{1}
          }

   \institute{Faculty of Science, Masaryk University, Department of Theoretical Physics and Astrophysics,  Kotl\'{a}\v{r}sk\'{a} 2, 611\,37 Brno,  Czech Republic\\
     \email{mikulas@physics.muni.cz, epaunzen@physics.muni.cz}
         \and
            National Astronomical Research Institute of Thailand, 260  Moo 4, Donkaew, Maerim, Chiangmai, Thailand 50180\\
            \email{eugene@narit.or.th}
         \and
        Bundesdeutsche Arbeitsgemeinschaft f\"{u}r Ver\"{a}nderliche Sterne e.V. (BAV), Berlin, Germany
        \and
        American Association of Variable Star Observers (AAVSO), Cambridge, USA
        \and
        Institut de Physique, Laboratoire d'astrophysique, Ecole Polytechnique F\'ed\'erale de Lausanne (EPFL), Observatoire de Sauverny, CH-1290 Versoix, Switzerland
             }

   \date{Received ; accepted }

 
  \abstract
  {CP2 stars show periodic photometric, spectroscopic, and magnetic variations with the rotational period. They are generally slow rotators, with rotational periods exceeding half a day, except for the late B-type star HD\,60431, which has an unusually short rotational period of 0.4755 days. As slow rotation is deemed a necessary criterion for the establishment of chemical peculiarities, this characteristic renders HD\,60431 a special object that might offer valuable insight into, and constraints on, the formation and evolution of CP2 stars.}
   {Our study aims at analysing the light variability, deriving atmospheric abundances, and determining detailed physical parameters of \hd\ to confirm its status as the CP2 star with the shortest known rotational period, with special emphasis on the rotational period evolution.}
   {Photometric indices and high-resolution spectroscopy were employed to derive physical parameters, evolutionary status, and atmospheric abundances of our target star. A light variability study was carried out using combined sets of photometric data from ground- and space-based facilities. A circularly polarised spectrum was employed to check the presence of a longitudinal magnetic field in the star.}
   {With an age of only 10 Myr, an effective temperature of $\Teff=13\,000\pm300$\,K, surface gravity log\,$g$\,=\,$4.10\pm0.10$, radius $R=1.97\pm0.09\,\Rs$, and mass $M=3.1\pm0.1\,\Ms$, \hd\ is situated close to the zero-age main sequence and a member of the open cluster NGC 2547 in the Vela OB2 complex. We confirm its status as a classical late B-type CP2 star showing strong overabundances of Mg (1.8~dex), Si (1.9~dex), Ca (1.6~dex), Ti (2.2~dex), and Fe (1.8~dex). No conclusive evidence for the presence of a strong magnetic field was found in the available spectroscopic data. The light curve of \hd\ has remained constant over the last four decades. The available photometric time series data confirm the short rotational period and indicate a slight secular increase of the rotational period of $\dot{P}=2.36(19)\times10^{-10}=7.5\,(6)\,\textrm{ms yr}^{-1}$. The following quadratic ephemeris has been derived: HJD$_\mathrm{min}(E)=2\,459\,212.969\,35+0\fd475\,516\,64\,E+5\fd62\times10^{-11}E^2$.}
   {HD\,60431 is indeed the CP2 star with the shortest known rotational period. Theory needs to explain the establishment and maintenance of chemical peculiarities in such a young and fast rotating object. Our results furthermore raise the question whether period variability  on timescales significantly shorter than stellar evolution is inherent to all magnetic chemically peculiar stars.}
   
   

   \maketitle
%

\section{Introduction}\label{intro}
About 10\,\% of upper main-sequence stars (spectral types from about early B to early F) show distinct spectral peculiarities indicative of peculiar surface abundances. These are the chemically peculiar (CP) stars, which are subdivided into several subgroups, such as the metallic-line (Am/CP1) stars, the Bp/Ap (CP2) stars, the HgMn (CP3) stars, and the He-peculiar stars \citep{preston74}. In general, the observed peculiar abundances are attributed to the processes of atomic diffusion, that is, the interplay between selective radiative levitation and gravitational settling taking place in the calm outer layers of slowly rotating stars \citep{michaud70,michaud81,richer00}.

\hd\,=\,V343\,Puppis, which is in the focus of the present investigation, belongs to the group of CP2 stars. These objects are characterised by overabundances of elements such as Si, Cr, Sr, or Eu, and possess globally organised magnetic fields with strengths of up to several tens of kG \citep{auriere07}. They show spectroscopic and photometric variability with the rotation period, which is generally of the order of several days, although objects with periods of years or even centuries are known \citep{rm09,mathys20}.

\hd\ was recognised as a CP2 star by \citet{bidelman73}, who identified Si and Mg peculiarities. \citet{north88} derived the unusually short photometric period of $P=0\fd475\,518(59)$~d and estimated an equatorial velocity of $V_{\mathrm{eq}}=245$\,\kms\ assuming a radius of $R$\,=\,$2.3$\,R$_\odot$. As rapid rotation induces meridional circulation that counteracts the diffusion processes, \hd\ poses a challenge to the atomic diffusion scenario. Although the effects of meridional circulation were described more than half a century ago \citep{michaud70,watson70, watson71}, they have been discussed in a rather qualitative way. A more quantitative study, such as was presented by \citet{quievy09} on meridional circulation in Horizontal Branch stars, is still lacking for CP2 stars. In this context, \hd\ may turn out to be a keystone in the understanding of the processes taking place in these objects.

Here, we present an investigation of \hd\ using newly acquired and archival photometric and spectroscopic observations, with the aim of deriving physical parameters and studying its chemical composition and period behaviour.

\section{Astrophysical parameters}\label{physta}
%
\subsection{Physical properties and evolutionary status}

Photometric calibrations from \citet{1993A&A...268..653N} and \citet{1997A&AS..122...51K} were applied to the Str\"omgren \citep{2015A&A...580A..23P} and Geneva photometric indices \citep{2022A&A...661A..89P} listed in the {\bf G}eneral {\bf C}atalogue of {\bf P}hotometric {\bf D}ata~\citep{1997A&AS..124..349M}\footnote{\url{https://gcpd.physics.muni.cz/}}. This provides initial estimates of the effective temperature and surface gravity of our target star. From Str\"omgren photometry, we derive a temperature of $T[u-b]=13\,280\pm270$\,K and a surface gravity estimate of $\log g=4.53\pm0.13$, while Geneva indices provide a hotter temperature of $T_\textrm{XY}=14\,200\pm400$\,K and $\log g=4.10\pm0.10$. 
 
The difference of about 1000\,K between the temperature estimates derived from the Str\"omgren and Geneva indices may be accounted for by the well-known fact that CP2 stars are characterised by a peculiar distribution of the overall stellar flux  \citep[e.g.][]{1973ApJ...185..577L,1974A&A....32..237L,1978A&A....63..155J,1975ApJ...195..397A,1981A&A....97..359N,stigler14}. Therefore, we employed the temperature domain corrections for CP stars developed by \citet{netopil08} and derive corrected effective temperatures of T$_{CP}$\,=\,$12\,550\pm850$\,K and T$_{CP}$\,=\,$12\,990\pm400$\,K from the Str\"omgren and Geneva photometric data sets, respectively.

From a fit to the observed hydrogen line profiles (cf. Section \ref{spect}), we derive $\Teff=13\,000\pm300$\,K, $\log g = 4.1\pm0.1$, and a projected rotational velocity value of $\vsini=190\pm20\,\kms$. Given~the peculiar composition of the star and the effects this may have on the observed colours as well as the fact that the hydrogen line profile is usually the best indicator of effective temperature in CP2 stars \citep{gray09}\footnote{ This is primarily the case for cool CP2 stars; for hotter CP2 stars such as \hd, the hydrogen lines are more sensitive to surface gravity. Nevertheless, simultaneous modelling of several hydrogen lines can still provide good $T_\mathrm{eff}$ estimates for these objects.}, we have adopted the values derived from the hydrogen line profile fitting as the most reliable parameters for further discussion. The temperature value derived in this way agrees very well with the corrected temperature values derived following \citet{netopil08}.

To estimate mass, radius, and age, we employed the Stellar Isochrone Fitting Tool\footnote{\url{https://github.com/Johaney-s/StIFT}}, which builds on the methodology of \citet{2010MNRAS.401..695M} in estimating the mass, age, radius, and evolutionary phase of a star according to its effective temperature and luminosity as calculated on the basis of $Gaia$ EDR3/DR3 data \citep{2021A&A...649A...1G}. The distance $d = 407.0^{+4.7}_{-5.0}$\,pc taken from \citet{2021AJ....161..147B} and the Bolometric Correction derived for CP stars by \citet{netopil08} were directly transformed into the luminosity (log\,$L/L_{\odot}$\,=\,$1.995\pm0.032$). The extinction in this region and distance from the Sun can be neglected which is supported by the standard dereddening routine in the Str\"omgren
photometric system \citep{1993A&A...268..653N}.
 
The tool automatically searches for similar data in models based on evolutionary tracks and selects the four grid points closest to the input value in the Hertzsprung-Russell diagram. From these grid points, the output parameters are obtained by repetitive linear interpolation. For the error estimation, the program uses the errors of the input parameters and a full statistical Monte-Carlo analysis. Assuming [Z]\,=\,0.014\footnote{The chosen [Z] value corresponds to solar metallicity. As described in the following subsection, \hd\ is most likely a member of NGC~2547, for which \citet{2020A&A...634A..34B} derive an average $[Fe/H] = 0.10\pm 0.01$ from the analysis of two members, which is slightly higher than solar. The corresponding shift on the HR diagram, however, is no larger than the error bars on $T_\mathrm{eff}$ and $L$ (see e.g. Fig.~15 of \citet{2012MNRAS.427..127B}).} \citep[cf.][]{2016A&A...585A.150N} and employing the isochrones of \citet{2012MNRAS.427..127B}, we obtain a radius of $R = 1.97\pm0.11$\,\Rs, a mass of  $M = 3.1\pm0.1$\,\Ms, a surface gravity of $\log g = 4.34\pm0.14$, and an age of about 10\,Myr, which indicates that the star is situated on the zero-age main sequence (ZAMS).

An overview over the astrophysical parameters derived by the different methods is presented in Table \ref{atmospheric_parameters}.

\begin{table}
\caption{Astrophysical parameters of HD\,60431, as derived by the indicated methods. For further analyses in this work, we adopted the values obtained through hydrogen line profile and isochrone fitting.}
\begin{tabular}{ll}
\hline
\hline
Method                               & Result    \\ \hline
Str{\"o}mgren $uvby\beta$ photometry & $T_{\mathrm{[u-b]}}$\,=\,$13\,280\pm270$\,K   									\\
                                     & log\,$g$\,=\,$4.53\pm0.13$                           \\
								 & $T_{\mathrm{CP}}$\,=\,$12\,550\pm850$\,K$^a$								 \\ \hline
Geneva photometry            & $T_{\mathrm{XY}}$\,=\,$14\,200\pm400$\,K 											 \\
                                     & log\,$g$\,=\,$4.10\pm0.10$
								 \\
                              & $T_{\mathrm{CP}}$\,=\,$12\,990\pm400$\,K$^a$
                                     \\ \hline
hydrogen line profile fitting        & $T_{\mathrm{eff}}$\,=\,$13\,000\pm300$\,K												\\ 
									 & log\,$g$\,=\,$4.10\pm0.10$ 											  \\
								 & $v\, \sin i =\,190\pm20$ \kms\										 \\ \hline
isochrone fitting$^b$        			 & $R$\,=\,$1.97\pm0.11$\,\Rs											 						 \\ 
									 & $M$\,=\,$3.1\pm0.1$\,\Ms												 			\\
									 & log\,$g$\,=\,$4.34\pm0.14$ 														 \\
									 & $t$\,=\,10\,Myr																 \\ \hline
\multicolumn{2}{l}{$^a$ Following \citet{netopil08}.} \\
\multicolumn{2}{l}{$^b$ Assuming [$Z$]\,=\,0.014.} \\
\hline
\hline
\end{tabular}

\label{atmospheric_parameters}
\end{table}

\subsection{\hd\ as a member of the Vela OB2 complex}\label{Vela}

\citet{2019A&A...621A.115C} investigated the young Vela OB2 association on the basis of $Gaia$ DR2 data \citep{2018A&A...616A...1G}, using proper motions, parallaxes (as a proxy for distance), and photometry to separate the various components of the Vela OB2 complex. This is a young stellar grouping (age between 10 and 30\,Myr) located in the direction of the constellations Vela and Puppis, at a distance of 350 to 400 pc from the Sun. It was initially identified by \cite{1914ApJ....40...43K}, who made a first attempt at finding the parallaxes of B-type stars
brighter than sixth magnitude. Using $Hipparcos$ data, \citet{1999AJ....117..354D} were able to perform a more detailed membership study, which led to the detection of a local group of pre-main sequence stars \citep{2000MNRAS.313L..23P} that is referred to as $\gamma$ Velorum cluster \citep{2016A&A...589A..70P}. The topology of the region is quite complex; there are a dozen of different regions interacting with each other to a certain extent. Most likely, this is the result of various star formation processes triggered by supernovae exploding 30\,Myr ago \citep{2019A&A...621A.115C}. 

\hd\ is located in the Vela OB2 complex. According to \citet{2019A&A...621A.115C}, there is a probability of 99\,\% that the star is a member of the open cluster NGC 2547. This agrees very well with the parameters and the age estimate of about 10\,Myr presented in Section \ref{physta}.


\section{Spectroscopic analysis}\label{spect}

\subsection{Spectroscopic observations and data reduction}\label{spec_obs}

For the spectroscopic analysis, we utilised an echelle spectrum of \hd\ collected in 2012 within the framework of the project ''Magnetism in Massive Stars (MiMeS)`` (programme ID 11BP14,  PI Gregg Wade) using the spectropolarimeter ESPaDOnS attached to the Canada-France-Hawaii Telescope (CFHT) in spectropolarimetric mode. The reduced 1D data were extracted from the scientific archive of the CFHT\footnote{\url{https://www.cadc-ccda.hia-iha.nrc-cnrc.gc.ca/en/cfht/}}. The spectrum covers the wavelength range  3\,690\:--\:10\,480\,\AA, with an average $\textrm{SNR}$ of 120 at 5\,500\,\AA. Spectral resolution $R$ is 65\,000. The intensity spectrum consists of four subexposures which were re-normalised separately order-by-order to the continuum level using routines from the astronomical data processing system IRAF\footnote{\url{https://github.com/iraf-community/iraf}} \citep{1999ascl.soft11002N}. We utilised a template spectrum synthesised with photometrically derived atmospheric parameters to reconstruct the continuum around the broad structures like hydrogen lines extending on multiple echelle orders. The average of the four subexposures trimmed to 3\,740\:--\:8\,000\,\AA\ was used in the subsequent analysis. Furthermore, a circularly polarised spectrum ($V$ Stokes) was employed to check the presence of a longitudinal magnetic field in the star.

The ESPaDOnS combined spectrum was obtained during the narrow phase interval of 0.015\,$\le$\,$\varphi$\,$\le$\,0.040. To increase phase coverage, an additional spectrum of \hd\ was taken with the echelle fibre-fed spectrograph MRES installed on the 2.4-m Thai National Telescope at Doi Inthanon (Thailand) in December 2021. The raw observational material was processed in a standardised way using the pipeline PyYAP, which was written in \textsc{Python} by one of the authors (ES) for this particular device\footnote{\url{https://github.com/ich-heisse-eugene/PyYAP}}. For the wavelength calibration, we used a ThAr spectrum recorded immediately after the star. After the application of the pipeline, the continuum level was adjusted using IRAF. The extracted and calibrated spectrum covers the wavelength region of 4\,200\:--\:7\,100\,\AA\ and has an average resolution of $R=17\,000$. The spectrum was employed to check the stability of the radial velocity and spectral variability of our target star. A fragment of the combined observational journal is shown in Table~\ref{tab:obslog}.
0
\begin{table}
\caption{Journal of spectroscopic observations used in this work. The upper five spectra were obtained with ESPaDOnS at the CFHT telescope, the bottom spectrum was obtained with MRES at the Thai National Telescope. Columns contain the following information: extracted file name, heliocentric Julian date of the middle time of the exposure (HJD), phase ($\varphi$) according to Eq. (\ref{varphi}),  exposure time ($T_\mathrm{exp}$), maximum signal-to-noise ratio of the data (SNR), type of data (Stokes $I$ or $V$).}
\tiny
\centering
\begin{tabular}{lccccc}
\hline\hline
Filename  &  HJD, 2450000+ & Phase $\varphi$ & $T_\mathrm{exp}$ [s] & SNR & Type \\
\hline
1513786i  &  5930.9635 & 0.015    & 300  & 146 & $I$  \\
1513787i  &  5930.9674 & 0.024    & 300  & 146 & $I$  \\
1513788i  &  5930.9714 & 0.032    & 300  & 148 & $I$  \\
1513789i  &  5930.9753 & 0.040    & 300  & 144 & $I$  \\
1513786p  &  5930.9694 & 0.028    & 1200 & 286 & $V$  \\
MRES      &  9565.2725 & 0.885    & 3600 & 170 & $I$  \\
\hline
\end{tabular}\label{tab:obslog}
\end{table}

\subsection{Astrophysical parameters and abundance analysis}\label{spec_par}

The fact that the star has been studied in the framework of the MiMeS survey and no results of magnetic measurements have been published so far suggests that the magnetic field is small, which allows us to neglect its influence in the process of spectrum modelling. We synthesised a spectrum assuming local thermodynamic equilibrium~(LTE) with the 'non-magnetic' code \textsc{Synth3}~\citep{2007pms..conf..109K, 2012ascl.soft12010K} that uses model atmospheres computed with \textsc{Atlas9}~\citep{2005MSAIS...8...14K, 2017ascl.soft10017K} and line lists extracted from the VALD database~\citep{2015PhyS...90e4005R}. 

\begin{figure*}
\begin{center}
\includegraphics[width=0.8\textwidth]{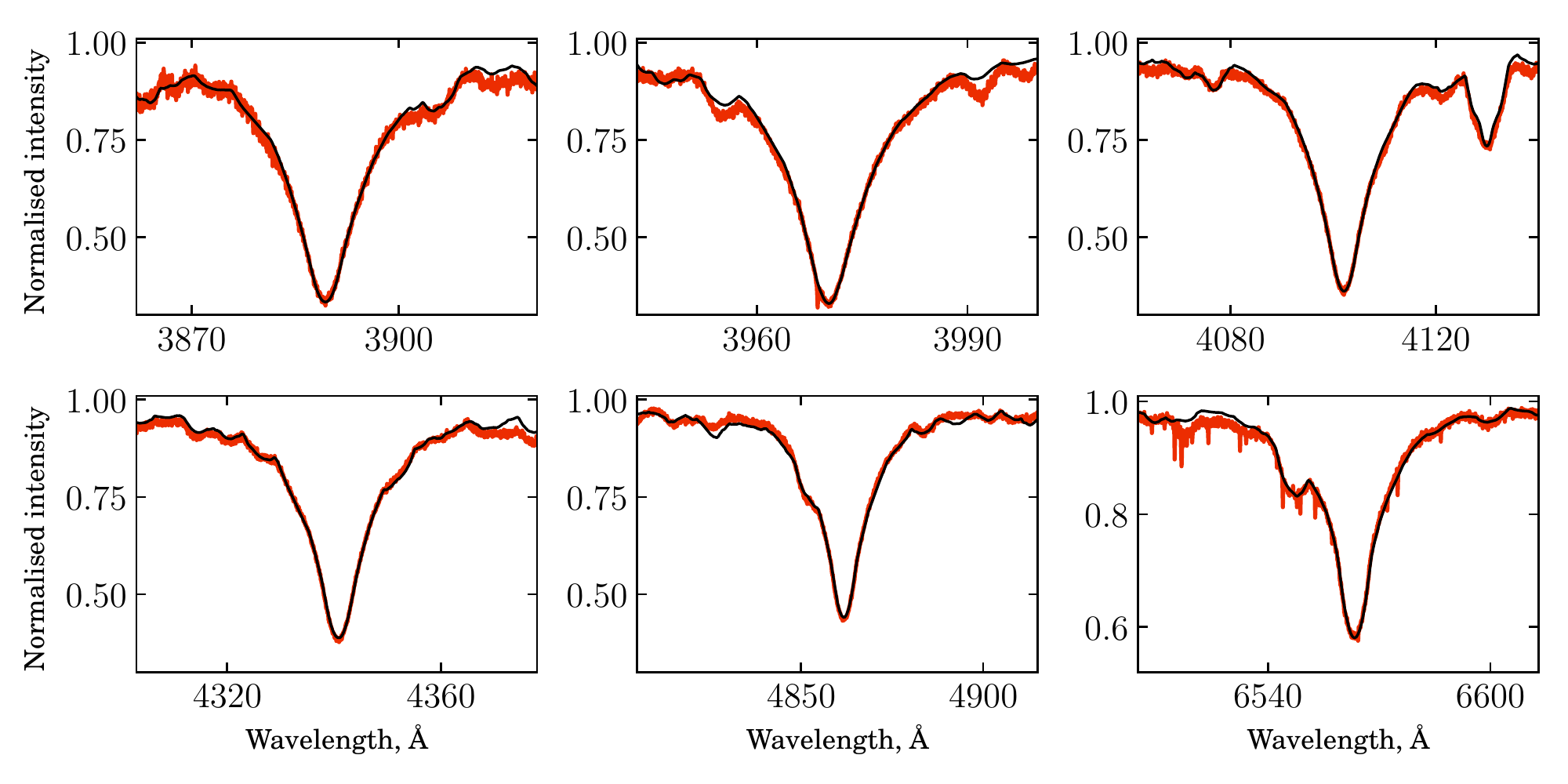}
\caption{Observed hydrogen line profiles (red) of \hd. The black lines indicate the fit used to derive the basic physical parameters.}
\label{parameters_fit}
\end{center}
\end{figure*}

From a simultaneous fit to the six observed hydrogen lines, we derive the effective temperature, surface gravity, and projected rotational velocity values presented in Section \ref{physta} ($\Teff=13\,000\pm300$\,K, $\log g = 4.1\pm0.1$, $\vsini=190\pm20\,\kms$). The corresponding fit is shown in Fig.\,\ref{parameters_fit}. It is important to note that, for a correct description of the observed level of the continuum, we had to significantly increase the abundances (and thus the number of lines) of most of the chemical elements involved in the peculiar abundance patterns of CP2 stars in our requests to VALD.

Chemical anomalies of the SiMg type were initially reported for \hd\ in the \cite{bidelman73} catalogue. Lines of Si and Mg are conspicuous in the observed spectrum, but numerous strong lines of iron are also present in blends with various elements.

\begin{figure*}
\begin{center}
\includegraphics[width=0.8\textwidth]{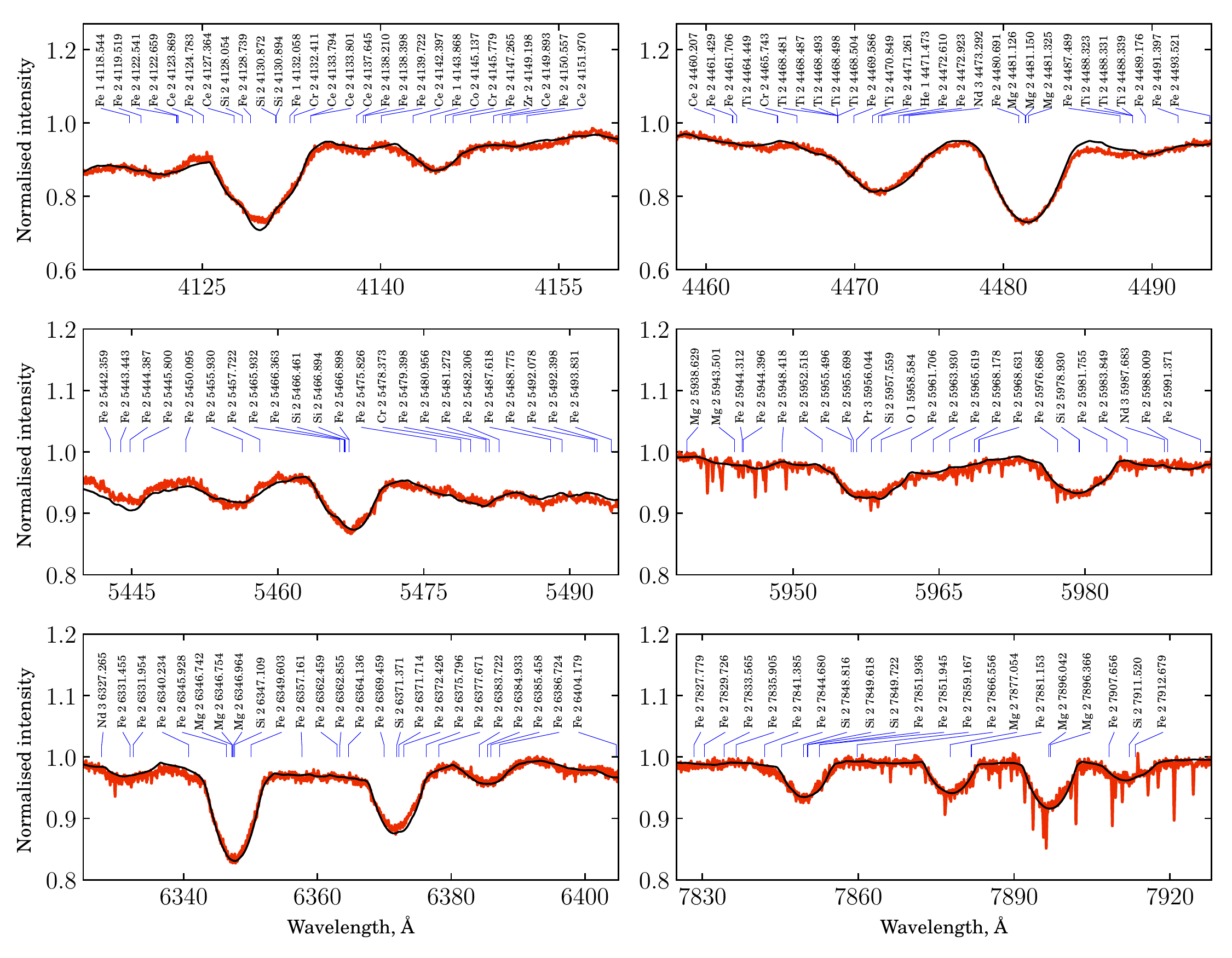}
\caption{Sample of spectral regions used for the evaluation of abundances. Red and black lines indicate the observed spectrum and the model fit, respectively.}
\label{spec_samples}
\end{center}
\end{figure*}

Adjusting the abundance of the main contributing elements and ions within the wavelength range of 4\,075\:--\:7\,990\,\AA, we simultaneously fitted 19 partially overlapping spectral regions rich in lines (Figure~\ref{spec_samples}). We found that Si and Mg are overabundant by 1.92 dex and 1.83 dex, respectively. Fe, the third most noticeable contributor, is enhanced by 1.77 dex with respect to the solar abundance.

In addition, several lines of He, O, and Ca were identified in the spectrum. He is represented by at least the He\,\textsc{i} 4\,471\,\AA\ and 5\,876\,\AA\ lines and, as deduced from an analysis of the blends with significant He contributions, has nearly solar abundance. This result is intriguing, because He is generally found deficient in late-B type CP2 stars, especially in young ones \citep{2014A&A...561A.147B}. \hd\ forms a noteworthy exception. Its comparably high He abundance might be linked to its fast rotation, and we may here have an intermediate case between the magnetic He-rich stars, which generally have a spectral type close to B2 \citep[see e.g.][]{2018A&A...618L...2J}, and the magnetic CP2 stars, which are generally He-weak. More He abundance determinations in rapidly rotating, late-B type CP2 stars are needed to clarify this point.

The observed peculiarly enhanced He\,\textsc{i} 5876\,\AA\ line (cf. Table \ref{tab:abn}) can be explained following the non-LTE approach~\citep{2018AstL...44..621K}, which is also required to account for the observed depth of the O\,\textsc{iii} 7771-5\,\AA\ triplet. Considering the derived oxygen excess of $\Delta\log \varepsilon=+2$ and the values of the non-LTE correction published by \citet{2016PASJ...68...32T}, we expect that the real abundance of O in \hd\ is close to solar. 

Ca is most prominent in the H and K Ca\,\textsc{ii} lines, from which only the latter line is suitable for modelling. The corresponding Ca\,\textsc{ii} K 3933\,\AA\ line indicates a Ca excess of 1.63 dex. However, the derived value does not take into account the influence of non-LTE effects and may suffer from missing weak blended lines. It is noteworthy that in addition to the broad stellar Ca lines, the observed spectra also contain sharp Ca components of interstellar origin.

The derived abundances of several other elements are reported in the overview provided in Table~\ref{tab:abn}. As these values have been derived mostly from modelling weak lines or complex blends, they should be regarded as providing general abundance trends only. The accuracy of the derived abundances depends mainly on the uncertainties of the physical parameters, the continuum normalisation, and the reliable identification of blends. In some cases, non-LTE effects require a separate modelling and were not included in the evaluation of errors. In this manner, we estimate that the typical error of the individual abundance measurements amounts to 0.20 dex for Mg, Si, and Fe and to 0.30 dex for He and Ti. Less accurate results in Table~\ref{tab:abn} are identified by a  colon (:).

To check for possible effects of a weak magnetic field (see Sec. \ref{spec_par}) on the derived abundances, we synthesised a spectrum of a star with the same physical parameters and a dipolar magnetic field of \bd\ = 1\,kG, using the code \textsc{SynthMag} written by \cite{2007pms..conf..109K}. Although the presence of the magnetic field visibly affects some individual lines, these effects are well contained within the errors. We therefore conclude that the influence of the magnetic field on the derived abundances is negligible.

\begin{table}
\caption{Derived chemical abundances of \hd. The solar abundance values have been compiled from \citet{2022A&A...661A.140M} (O, Mg, Si, Ca, and Fe) and \citet{2021A&A...653A.141A} (all other elements). $\Delta\log \varepsilon$ is the abundance relative to solar, defined by $\Delta\log \varepsilon=\log\left(\frac{N/N_\mathrm{tot}}{(N/N_\mathrm{tot})_{Sun}}\right)$. A colon indicates an uncertain value.}
\tiny
\centering
\begin{tabular}{lccc}
\hline\hline
 Element  &  $\log N/N_\mathrm{tot}$  & $(\log N/N_\mathrm{tot})_\mathrm{Sun}$ & $\Delta\log \varepsilon$ \\
\hline
He    &  $-1.05$ ($-0.80$/$\lambda5876$\,\AA)  &  $-1.12$ &  0.07\\
O     &  $-1.30$ (:)                           &  $-3.34$ &  2.04 (:)\\
Na    &  $-5.00$ (:)                           &  $-5.81$ &  0.81 (:)\\
Mg    &  $-2.65$                               &  $-4.48$ &  1.83\\
Si    &  $-2.60$                               &  $-4.52$ &  1.92\\
Ca    &  $-4.10$ (:)                           &  $-5.73$ &  1.63 (:)\\
Ti    &  $-4.90$                               &  $-7.06$ &  2.16\\
Cr    &  $-4.70$ (:)                           &  $-6.41$ &  1.71 (:)\\
Fe    &  $-2.75$                               &  $-4.52$ &  1.77\\
Pr    &  $-8.00$ (:)                           &  $-11.28$&  3.28 (:)\\
Nd    &  $-8.00$ (:)                           &  $-10.61$&  2.61 (:)\\
Tb    &  $-7.60$ (:)                           &  $-11.72$&  4.12 (:)\\
\hline
\end{tabular}\label{tab:abn}
\end{table}

In summary, our results indicate that, in accordance with the literature, \hd\ is a highly peculiar late B-type CP2 star. Besides significant overabundances of Mg, Si, Fe, Cr, and Ti, we find a strong excess of many other elements, including three lanthanides. The classification as a CP2 star is further corroborated by the presence of a conspicuous flux depression at around 5\,200\,\AA, which has been shown to be a characteristic of magnetic chemically peculiar stars \citep{1969ApJ...157L..59K, 1975ApJ...195..397A,1976A&A....51..223M, 1979AJ.....84..857A,2006MNRAS.372.1804K,2020A&A...640A..40H}. This flux depression is detected in the Geneva photometric system through the photometric parameters $\Delta(V1-G)$ \citep{1978A&A....69..285H} and $Z$
\citep{1979A&A....78..305C}. From the average Geneva colors of \hd\ taken from \citet{1989A&AS...78..469R} and
the definitions of these parameters, we obtain $\Delta(V1-G)=0.072$ and $Z=-0.092$. As normal stars have $\Delta(V1-G)\simeq-0.005$ and $Z\simeq 0.00$, we see that \hd\ 
has extreme values of both parameters ($Z$ is negative for CP2 stars). While the depth of the depression 
depends on $T_\mathrm{eff}$, it achieves maximum strength at around 12\,000\:--\:13\,000~K, just at the effective temperature of \hd.

\subsection{LSD profiles and magnetic field}\label{spec_par}

The Least-Squares Deconvolution (LSD; \citealt{2010A&A...524A...5K}) profiles of \hd\ were derived from the echelle-spectra obtained in 2012 and 2021 and listed in Table \ref{tab:obslog}. To extract the averaged profiles, we used a mask based on a line list from VALD that represents the physical parameters of a typical CP2 star of temperature type B9. The spectral range and resolution of the two spectra were adjusted to each other.

\begin{figure}
\begin{center}
\includegraphics[width=0.40\textwidth]{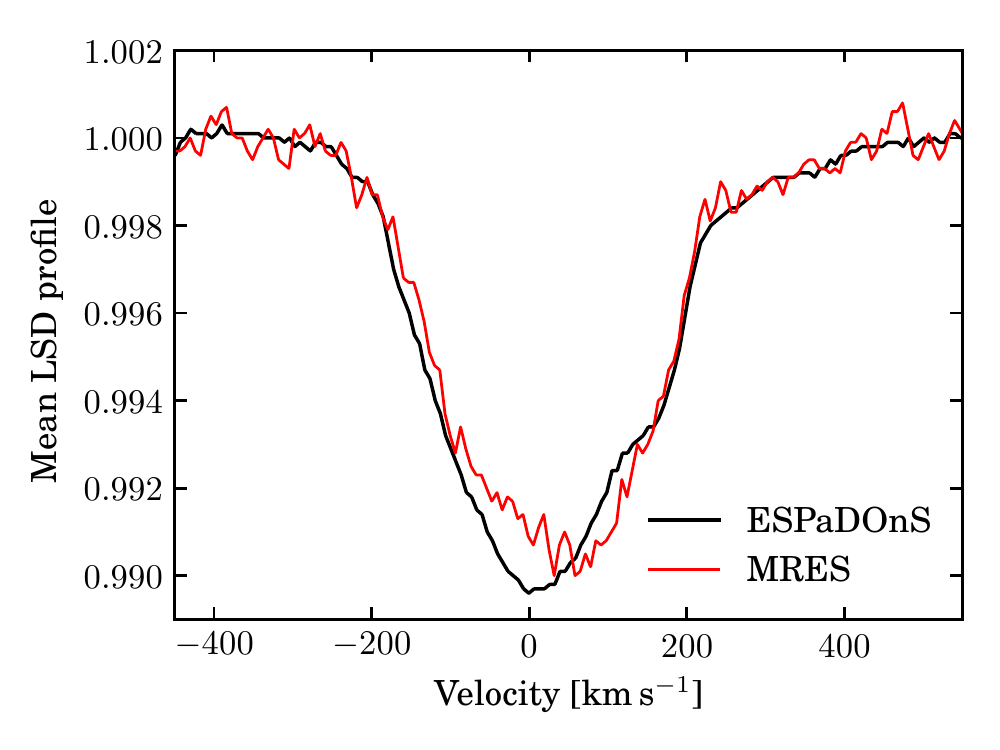}
\caption{Mean LSD-profiles derived from echelle-spectra obtained with ESPaDOnS and MRES in 2012 and 2021.} 
\label{fig:lsd}
\end{center}
\end{figure}

The LSD profiles have variable shapes, which are most likely caused by the presence of abundance spots ~(Fig. \ref{fig:lsd}).
Distinctive spot signatures were found for Si and Fe after adjusting the mask to the observed spectrum. This finding agrees with the expectations for CP2 stars and with the observed rotational modulation in our target star.

\begin{figure}
\begin{center}
\includegraphics[width=0.4\textwidth]{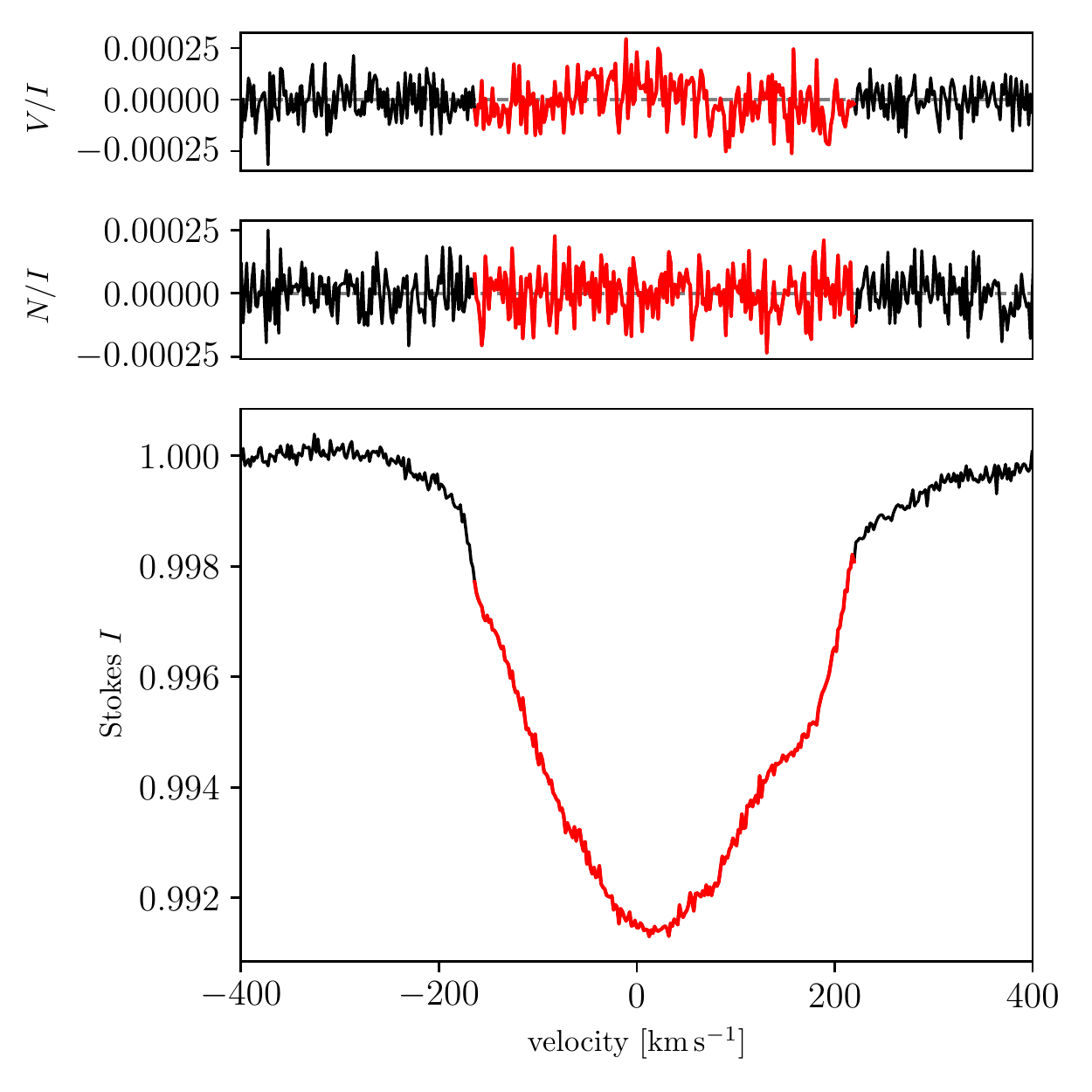}
\caption{Mean LSD-profiles used for measuring of the longitudinal magnetic field. Red part of profiles shows the integration limits.}
\label{fig:mlsd}
\end{center}
\end{figure}

The radial velocity $V_\mathrm{r}$ as measured from the LSD-profiles amounts to 18.4 km\,s$^{-1}$ in 2012 and 23.2 km\,s$^{-1}$ in 2021. The difference in $V_\mathrm{r}$ can be explained by the variable core and falls well within the measurement error of 5--8 km\,s$^{-1}$. From the available data, no meaningful conclusion can be drawn about a possible multiplicity of \hd.

As has been pointed out above, \hd\ has been studied in the framework of the MiMeS survey and no results of magnetic measurements have been published. Along with the intensity spectra, we therefore extracted one circularly polarised spectrum from the CFHT archive. To check the presence of a significant longitudinal magnetic field, we analysed the spectropolarimetric data based on the LSD technique described in \citet{1997MNRAS.291..658D} and \citet{2010A&A...524A...5K}. From the corresponding LSD profiles, in the velocity space from $-165$ to 223 km\,s$^{-1}$, we measure a mean longitudinal magnetic field of $\langle B_z \rangle = 270\pm150$\,G.

Despite the fact that a typical Zeeman signature is visible in the mean $V$ Stokes profile, we cannot assert that the star is truly magnetic. From a statistical analysis of the signal, we derive a false alarm probability of 0.99. This result implies that the magnetic field of \hd\ is either very weak or could not be detected because of insufficient data quality or an observation at an inappropriate phase of the magnetic curve (see discussion in Sect.\,\ref{LC_inter}). In summary, our analysis provides evidence that under the assumption of a dipolar field, \hd\ has a weak magnetic field of the order of 1 kG on the pole, which is, however, not conclusive and needs to be verified by additional studies.

\section{Photometric analysis} \label{perana}
This section contains a description of the observed light curves of \hd\ in the wavelength range 3\,400\:--\:8\,000\,\AA\ and an analysis of the period evolution in the last decades. For the latter purpose, we relied on our own methods of phenomenological modelling of light curves or their segments \citep[][and references therein]{Mik16} and their solution by robust regression \citep{Mik20b}.
\subsection{Observations}
Our study is based on diverse data sources. We took into account 357 individual observations in seven bands of the Geneva system (see Table\,\ref{ampl}), which were obtained using the Swiss telescope at the European Southern Observatory, La Silla (Chile) in 1981-88 and published by \citet{north88}. Furthermore, 667 measurements in the $V$ band were gleaned from the archives of the third phase of the All Sky Automated Survey (ASAS-3; \citealt{ASAS}), which cover the period 2000-09 and were divided into two parts: ASAS\,3-I and ASAS\,3-II (see Table\,\ref{oc}). The most accurate information on \hd, however, is represented by four data sets obtained by the Transiting Exoplanet Survey Satellite (TESS) \citep{ricker14,ricker15}, which contain a total of 8\,911 measurements collected in 2019 and 2021 (sectors 7, 8, 34, and 35).

\subsection{Multicolor light curve modelling}
According to \citet{north88}, \hd\ shows a fairly large amplitude in the $U$ band of almost 0.1 mag, while the amplitude in the $V$ band hardly reaches 0.04 mag (see Table\,\ref{ampl}); all colours were shown to vary in phase.

\begin{figure}
\begin{center}
\includegraphics[width=0.40\textwidth]{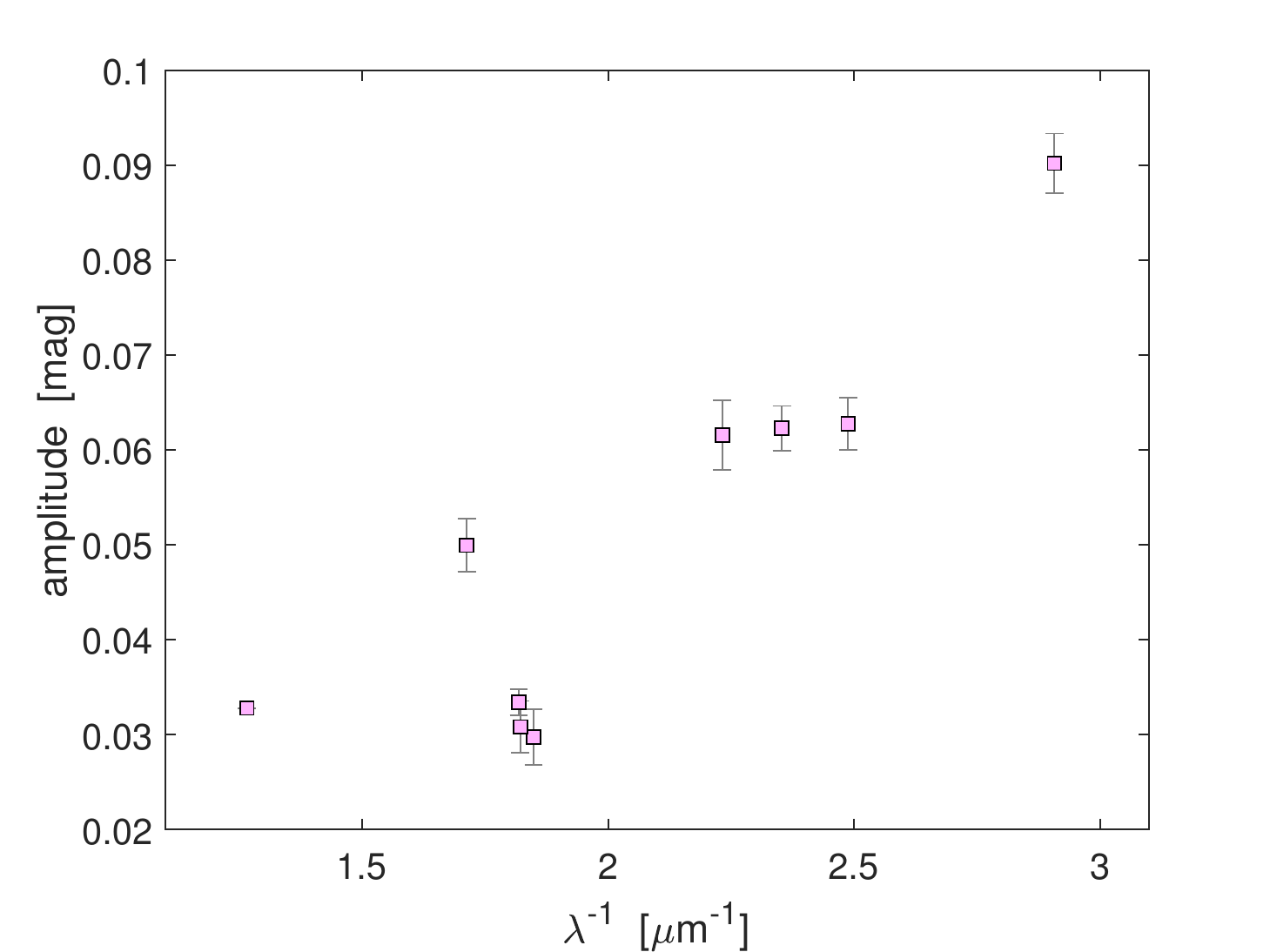}
\caption{Dependence of effective amplitudes (as defined in \citealt{Huemmerich18}) of light curves in the $[U],B1,[B],B2,V1,[V],V,G$, and $T$ bands on the reciprocal wavelengths (see Table\,\ref{ampl}). }
\label{amplitudy}
\end{center}
\end{figure}

\begin{figure}
\begin{center}
\includegraphics[width=0.55\textwidth]{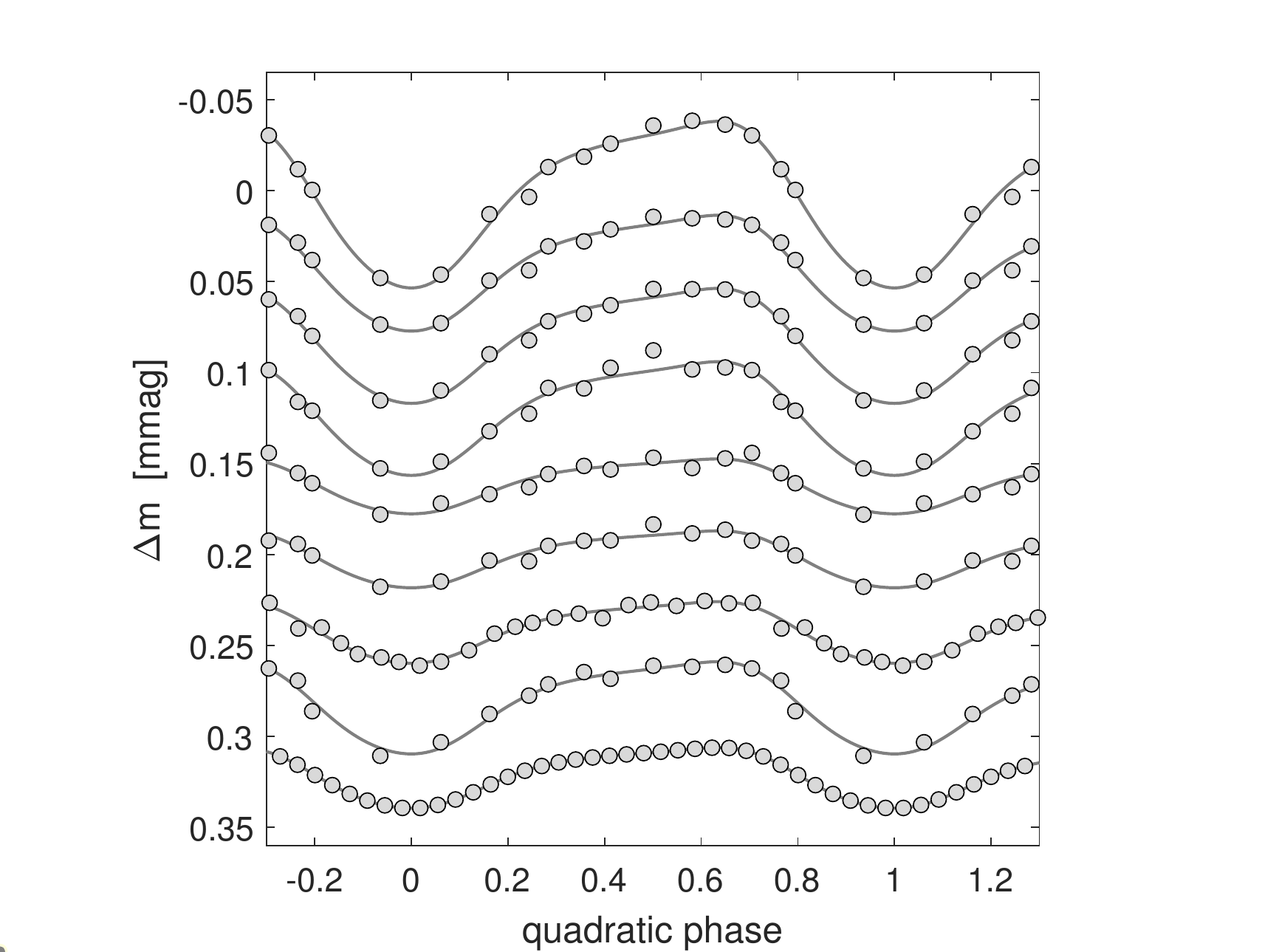}
\caption{Light curves of \hd\ in $[U],B1,[B],B2,V1,[V],V,G$, and $T$ (see Table\,\ref{ampl}). The mean light curves are defined by the filled dots, which represent 'normal points' as defined in Appendix\,\ref{normalka}. The model light curves are represented by the solid lines.}
\label{krivky}
\end{center}
\end{figure}

To investigate the relationship of the phased light curves in different passbands quantitatively, we applied the technique of advanced principal component analysis \citep{Mik04} to the Geneva photometry of \hd. It was found that the light curves phased on the period of $P=0\fd475\,518$ \citep{north88} display only one significant principal component. All other components can be neglected without a deterioration in the accuracy of the fit. This finding also applies to the ASAS-3 and TESS light curves (see Fig.\,\ref{krivky}).

While CP2 stars may show very different light curves at different wavelengths, it has been shown by \citet{Mik04,Mik08a} that similar light curve patterns in different passbands are commonly observed in these objects. Obviously, in these stars, only one dominant mechanism is responsible for the redistribution of the flux in the spectrum. All the light curves $F(\lambda,\varphi)$ of \hd\ can be satisfactorily approximated by the simple model:
\begin{equation}
F_{k}(\lambda,\varphi)=A(\lambda)\,B(\varphi,\vec{\alpha}) + m_{0k},
\label{model}
\end{equation}
where $A$(\lam) is the effective semiamplitude of the light curve in the individual photometric passbands, $B(\varphi,\vec{\alpha})$ is the normalised periodic function of the phase function $\vartheta$. In this case, we define $B(\varphi,\vec{\alpha})$ as a linear combination of nine simple mutually orthonormal periodic functions, of which five are symmetric and four antisymmetric functions to the phase $\varphi = 0$, which describes the eight independent parameters forming the alpha vector $\vec\alpha$. More details are given in Appendix \ref{LC model}.

\begin{table}
\caption{Photometric bands of the observations used in this study (cf. Figs.\,\ref{amplitudy} and \ref{krivky}). Brackets are used to distinguish the Geneva $[U]$, $[B]$, $[V]$ from the Johnson $U$, $B$, $V$ passbands, which are slightly different. $\sigma$ is the RMS scatter of the residuals after subtraction of the model light curve.}
\tiny
\centering
\begin{tabular}{cclrc}
\hline\hline
 Color & Wavelength &\ Eff. ampl. & Number& $\sigma$\\
 & [nm] &\ \ [mag]& & [mmag]\\
\hline
$[U]$       &   344  &   0.090(3)  &        51 &    8 \\
$B1$       &   402  &   0.063(3)   &        51 &    7 \\
$[B]$       &   425  &   0.062(2)  &        51 &    6 \\
  $B2$     &   448  &   0.062(4)   &        51 &    9 \\
 $V1$      &   541  &   0.030(3)   &        51 &    8 \\
 $[V]$     &   549  &   0.031(3)   &        51 &    7 \\
$V$        &   550  &   0.0334(14) &       667 &   13 \\
 $G$      &   584  &   0.050(3)    &        51 &    7 \\
 $T$      &  790   &   0.0327899(17)  &   7887 & 0.55 \\
 \hline
 \end{tabular}\label{ampl}
\end{table}

\subsection{Phase function model}
The phase function $\vartheta(t)=E+\varphi$ ($E$ is the epoch and $\varphi$ the common phase) and its inverse function  $\T(\vartheta)$ are related to an instantaneous period $P(t)$ at the time $t$ through simple differential equations with a boundary condition \citep[][]{Mik08b,Mik16}:
\begin{align}
&\frac{\rm d\vartheta}{{\mathrm d}t}=\frac{1}{P(t)} ;\quad \vartheta(t=M_0)=0;\quad
\Rightarrow\quad \vartheta(t)= \int_{M_0}^t \frac{\rm d \tau}{P(\tau)};\label{fazovka}\\
&\frac{\mathrm{d} \T(\vartheta)}{\rm d\vartheta}=P(\vartheta);\quad \T(0)=M_0;\quad
\T(\vartheta)= M_0+\int_0^\vartheta P(\zeta)\,\rm{d}\zeta;\label{inverze}
\end{align}

Using $\T(\vartheta)$, we can calculate the moments of the zero phases JD$_{\mathrm{min}}(E)=T(E)$, where $E$ is the epoch.

\subsubsection{Linear model solution}
If the period of the variable star is constant $(P(t)=P_1)$, the solutions of the equations (\ref{fazovka}) and (\ref{inverze}) are trivial:
\begin{equation}
\vartheta_1=(t-M_1)/P_1;\quad \Rightarrow \quad \mathrm{JD_{min}}_1(E)=M_{1}+P_1\,E,\label{linda}
\end{equation}
with only two parameters of the linear ephemeris: $P_1$ and $M_{1}$. The linear model of the phase function can be introduced into light curve model of \hd\, and all free parameters can be found simultaneously by robust regression, which suppresses the influence of outlying data points \citep[see e. g.][]{Mik20b}. Following this approach, we derived the following linear ephemeris:
 $M_{1}=2\,459\,219.151\,09(16)$ and $P_1=0\fd475\,515\,90(7)$, which roughly describes the course of the phase function $\vartheta(t)$ and its inverse $\T(\vartheta)$.

However, a detailed analysis indicated that the basic assumption of a constant period $P$ is not fulfilled. This can be demonstrated, for example, by the fact that the linear periods derived from different data sets are not compatible with each other. Omitting the TESS observations, we derive a period of $P_1=0\fd475\,515\,456(20)$, ($M_{1}=2\,449\, 538.609\,6(21)$). If we then omit the Geneva and ASAS-3 measurements, a period of $P_1=0\fd475\,516\,50(11)$, ($M_{1}=2\,459\,218.675\,57(16)$) is obtained. This can be interpreted as a lengthening of the period by $0.17\pm0.02$\,s over a time span of 26.5 years. The same result is derived from an analysis of the (O-C) values for individual groups of observations (see Fig.\,\ref{ocko}).

We therefore conclude that the period of \hd\ has not remained constant over the observed time span. The simplest model that adequately describes the observations is provided by the assumption that the period has increased linearly with time, so $\dot{P}>0$.

\subsubsection{Quadratic model solution}\label{kvamodel}

Assuming that the rate of change of the period length $\dot{P}$ has remained constant during the whole interval of the observations, we find that the phase function $\vartheta(t)$ is expressed by a quadratic polynomial, which we adjust to its orthogonal form:
\begin{align}
&\vartheta_1(t)=\zav{t-M_{1}}/P_1;\quad a_1=\overline{\vartheta_1^3}/\overline{\vartheta_1^2};\quad a_2=\overline{\vartheta_1^2};\nonumber \\ 
&\vartheta=\vartheta_1-\textstyle{\frac{1}{2}}\dot{P}\zav{\vartheta_1^2 -a_1\vartheta_1-a_2},\quad E=\mathrm{IP}(\vartheta),\ \varphi=\mathrm{FP}(\vartheta)\label{varphi}\\& P(\vartheta_1)=\zav{\frac{\mathrm d\vartheta}{\mathrm dt}}^{-1}=P_1\zav{\frac{\mathrm d\vartheta}{\mathrm d\vartheta_1}}^{-1} \doteq P_1+\dot{P}P_1\zav{\vartheta_1- \textstyle{\frac{1}{2}}a_1},\\
&\mathrm{JD}_{\rm{min}}(E)=M_{1}+P_1\,E+\textstyle{\frac{1}{2}}\dot{P}P_1
\zav{E^2-a_1E-a_2},\label{kvadra}
\end{align}
where function IP$(\vartheta)$ is the integer part of $\vartheta$, while FP$(\vartheta)$ denotes the fractional part of it. $P_1$ is the linear approximation to the period, $M_{1}$ is the time of light minimum near the weighted center of all observations, $\vartheta_1(t)$ is the linear phase function, $a_1,\ a_2$ are the quantities that characterise the distribution of the observations and their quality. $P(\vartheta_1)$ is a prediction of the instantaneous period at time $t=M_{1}+P_1\vartheta_1$, and JD$_{\rm{min}}(E)$ predicts the Julian date of the light minimum for epoch $E$.

Solving the model of the light variations, we find the following period model parameters:
$M_{1}=2\,459\,212.969\,37(17),\ P_1=0\fd475\,515\,63(5)$, $\Om_1=2\,\pi/P_1= 1.529\,330\,4\times10^{-4}$\,s$^{-1},\ \dot{P}=2.36(19)\times 10^{-10}=7.4(6)\,\textrm{ms yr}^{-1}$, and $\dot{P}/P_1=-\dot{\Om}/\Om_1=5.0(4)\times10^{-10}$\,d$^{-1}$. $a_1=-1.797\times 10^4$ and $a_2=3.527\times 10^{5}$.
The orthogonality of the model enables an easy estimate of the uncertainties of all quantities involved in the model parameters.

\begin{figure}
\begin{center}
\includegraphics[width=0.47\textwidth]{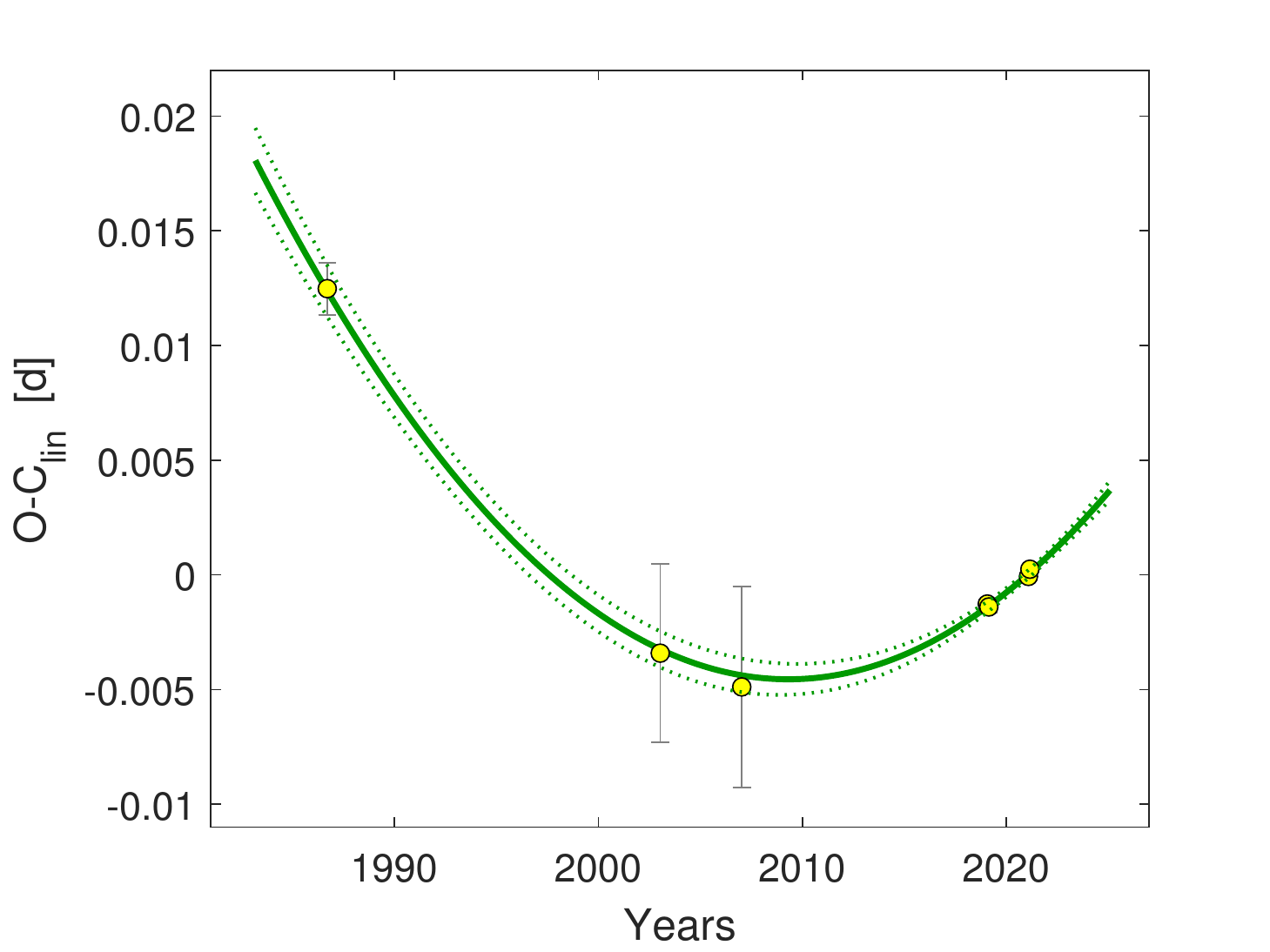}
\caption{Virtual O-C$_{\mathrm{lin}}$ diagram derived using relations in \citet{Mik12}, see also Table\,\ref{oc}. Full line denotes the model, dashed lines its 1-$\sigma$ uncertainty. }
\label{ocko}
\end{center} 
\end{figure}

\begin{table*}
\caption{Virtual O-C values \citep{Mik12} for individual groups of photometric data. $E_{\mathrm{mean}}$ is the mean epoch, HJD$_{\mathrm{min}}$ is the corresponding HJD moment of the light minimum, O-C$_{\mathrm{lin}}$, and O-C$_{\mathrm{sqrt}}$ are O-C values computed from linear (\ref{linda}) and quadratic (\ref{kvadra}) ephemerides, $\delta\,$HJD$_{\mathrm{min}}$ is the uncertainty on the minimum moment determination, all in days. $\sigma$ is the scatter of data in mmags around the fitted model curve, $N$ is the number of measurements.}
\centering
\begin{tabular}{lrllllcr}
\hline\hline
  System & $E_{\rm mean}$ & \quad HJD$_{\rm min}$ & O-C$_{\rm{lin}}$ &O-C$_{\rm {sqrt}}$ & $\delta\,$HJD$_{\rm min}$& $\sigma$ [mmag]& $N$\ \ \\
\hline
Geneva          &-26\,340 & 2\,446\,687.912\,3  &  \ 0.001\,3   &  \ 0.000\,1   &    0.001\,2    &  \ 7   &   357  \\
ASAS 3-I        &-13\,802 & 2\,452\,649.896\,1  & -0.003\,4   &  -0.000\,2   &    0.003\,9    & 12  &    299  \\
ASAS 3-II       & -10\,733 & 2\,454\,109.251\,3  & -0.004\,9   &  -0.000\,5   &    0.004\,4    & 14  &    368  \\
TESS, sector 7    &     -1\,490 & 2\,458\,504.448\,28 &  -0.001\,27  &  \ 0.000\,13  &    0.000\,21   &\ 1.2 &   1\,059 \\
TESS, sector 8    &     -1\,432 & 2\,458\,532.028\,28 & -0.001\,40  &  -0.000\,05   &   0.000\,26   &\ 1.3 &    867 \\
TESS, sector 34    &     60 & 2\,459\,241.500\,39 &  -0.000\,07  &   -0.000\,11  &    0.000\,05   &\ 0.5 &   3\,342 \\
TESS, sector 35    &     115 & 2\,459\,267.653\,86 & \ 0.000\,26  &  \ 0.000\,15   &   0.000\,05   &\ 0.5 &    2\,619
\\
\hline
\end{tabular}\label{oc}
\end{table*}
The times of the light minima for the observational subsets listed in Table\,\ref{oc} and the corresponding O-C diagram (Fig.\,\ref{ocko}) are visualisations of the model results according to the relations given in \citet{Mik12}.

Ignoring the uncertainties of the predicted quantities, the quadratic ephemeris can be represented in the more elegant form of
\begin{align}
& M_0=M_1-\textstyle{\frac{1}{2}}\,a_2\dot{P}\,P_1;\ \ P_0=P_1-\textstyle{\frac{1}{2}}\,a_1\dot{P}\,P_1;\ \ \displaystyle \vartheta_0=\frac{t-M_0}{P_0};\\
& \vartheta=\vartheta_0-\textstyle{\frac{1}{2}}\,\dot{P}\,\vartheta_0^2,\quad \displaystyle P(t)=\zav{\frac{\mathrm d\vartheta}{\mathrm dt}}^{-1}=P_0+\dot{P}\zav{t-M_0},\\
&\T=M_0+P_0\vartheta+\textstyle{\frac{1}{2}}\dot{P}P_0\vartheta^2,\ \ \mathrm{JD}_{\rm{min}}=M_0+P_0E+\textstyle{\frac{1}{2}}\dot{P}P_0E^2,\\
& \mathrm{JD}_{\rm{min}}(E)=2\,459\,212.969\,35+0\fd475\,516\,64\,E+5\fd62\times10^{-11}E^2,\nonumber
\end{align}
where $P_0,\ M_0$ and $\dot P$ are the coefficients of the standard Taylor quadratic decomposition at the time $t = M_0$. $P_0=0\fd475\,516\,64,\ M_0=2\,459\,212.969\,35
\doteq M_{1}$.

\begin{figure}
\begin{center}
\includegraphics[width=0.37\textwidth]{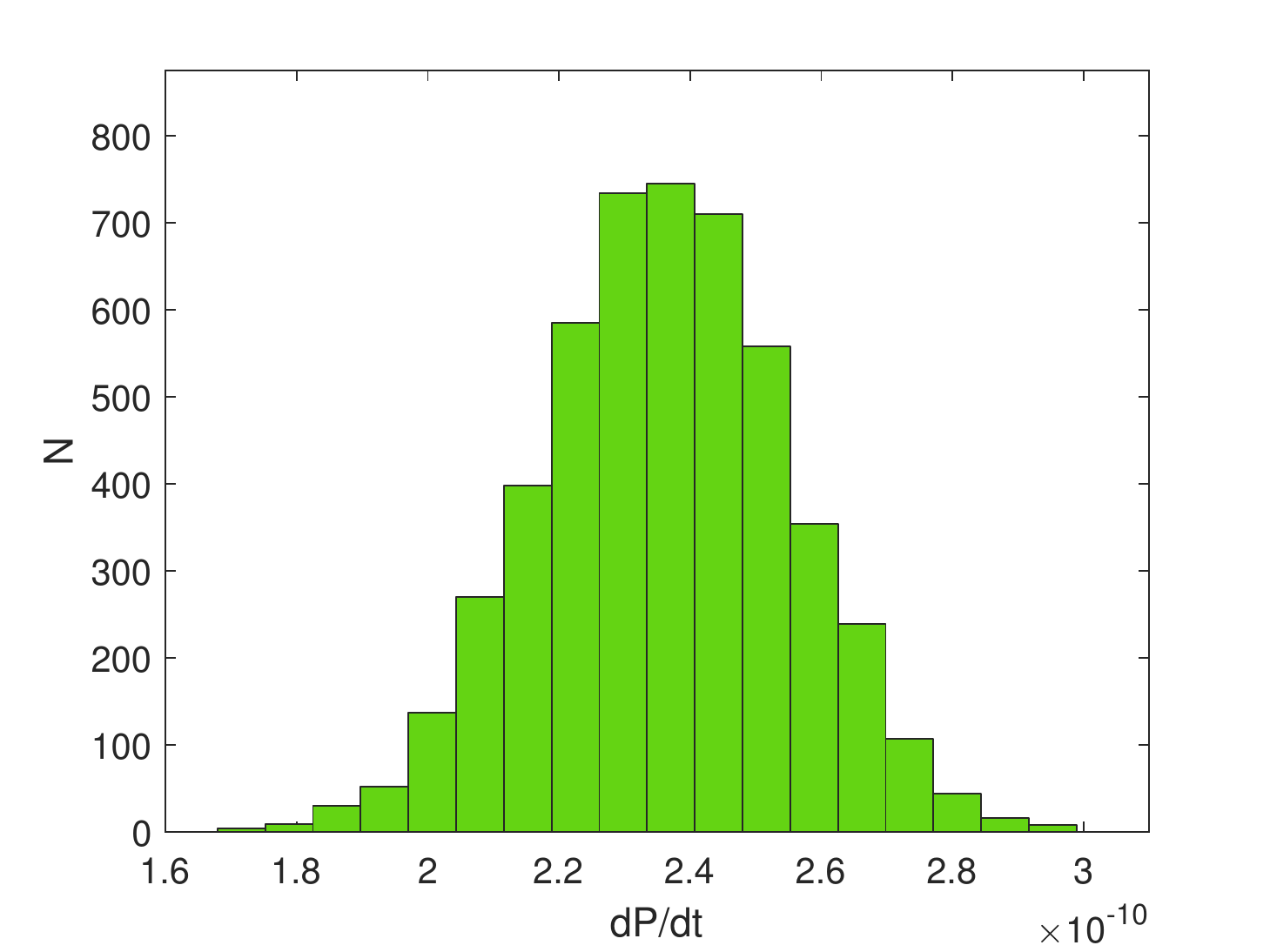}
\caption{Histogram of the results of 5\,000 bootstrap trials for $\dot{P}$ proves its non-zero value.}
\label{boot}
\end{center}
\end{figure}

To test the reality of the observed period change in \hd, we subjected the model calculations to several tests --- relying, in particular, on a bootstrap approach with $5\,000$ cycles --- and checked the robustness of the results to estimate the uncertainties. We found a perfect agreement of the results obtained with the above method and verified that the data variance entirely agrees with the Gaussian ideal (see Fig.\ref{boot}). 
In summary, the available evidence proves that the observed period change in \hd\ is real.

\begin{figure}
\begin{center}
\includegraphics[width=0.40\textwidth]{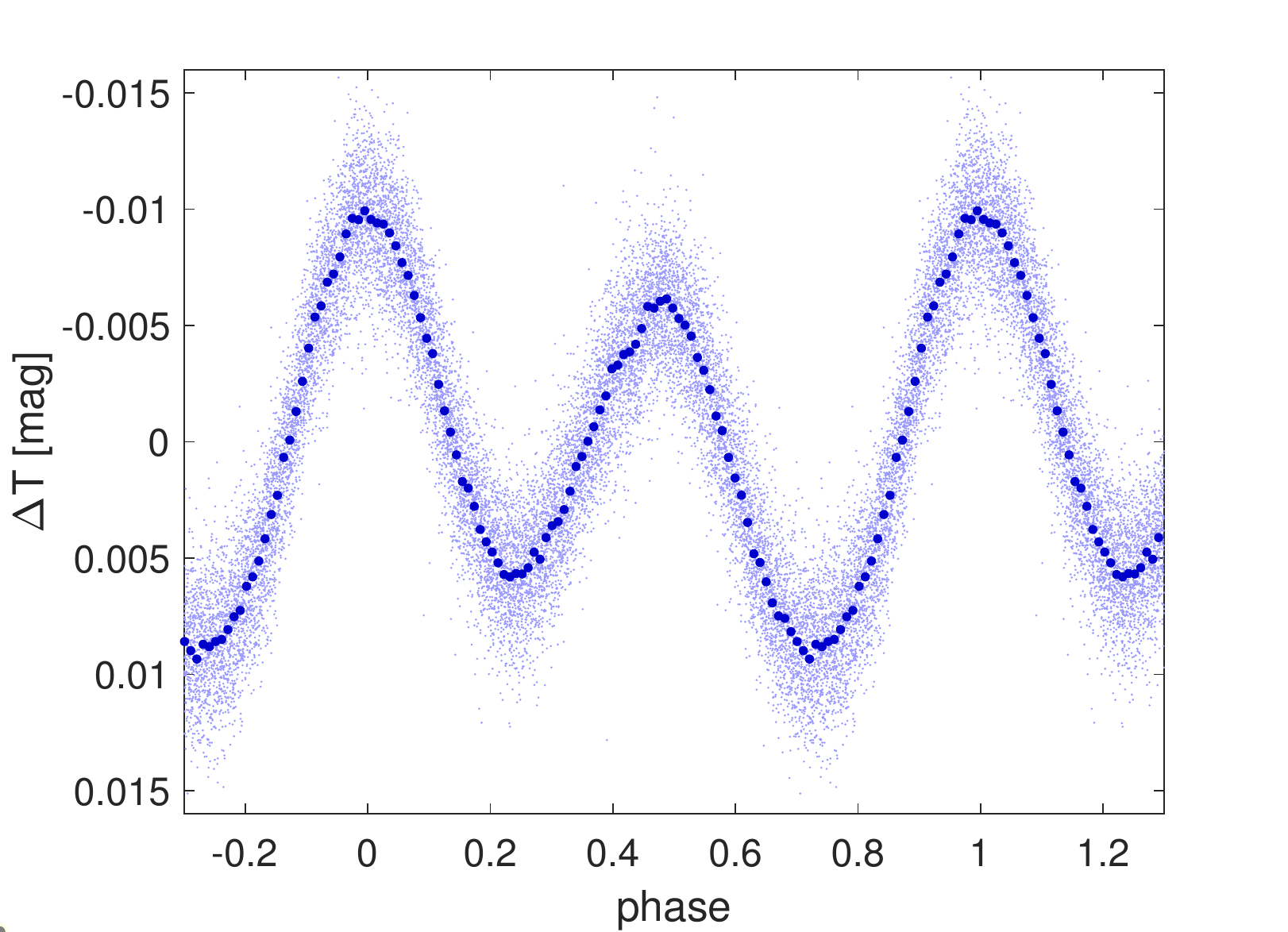}
\caption{Phased TESS light curve of HD\,98000, which clearly shows a double-wave light curve. Small blue dots denote detrended observations; large blue dots are 'normal points' representing the mean phased light curve (see Appendix\,\ref{normalka}).}
\label{hd98}
\end{center}
\end{figure}

\section{Discussion}

\subsection{Is \hd\ the CP2 star with the shortest known rotational period?}\label{98000}

Since the first period determination by \citet{north88}, \hd\ has been considered as the CP2 star with the shortest known rotational period. However, this perception has been challenged recently by \citet{huemmerich17}. Using photometric time-series data from ASAS-3, these authors derived a rotational period of $P=0\fd46566$ for the Si CP2 star HD\,98000, which -- if confirmed -- would represent the shortest observed rotational period among this group of objects. However, the authors cautioned that the amplitude of the variability is very small and that a twice longer rotation period could not be excluded on the basis of the available data, and called for additional observations to further investigate this matter.

HD\,98000 (TIC~82196009) was observed in Sector 10 by TESS for a duration of $\sim$26 days. The ultra-precise TESS data unambiguously show that, in contrast to the findings from the ground-based observations, HD\,98000 indeed shows a double-wave light curve (Fig.\,\ref{hd98}). Obviously, therefore, the period solution of \citet{huemmerich17} represents only half the true value. From an analysis of TESS data, we derive a rotational period of $P=0\fd93130(3)$. \hd, therefore, remains the CP2 star with the shortest known rotational period.





\subsection{Period variability}
We currently know of nine magnetic chemically peculiar stars, including \hd, that have shown changes of their rotational periods \citep{Mik21}. These objects form a very heterogeneous group, with some members showing increasing or decreasing periods as well as stars with more complex period behaviour. With an observed value of $\dot{P}=2.36(19)\times10^{-10}$, \hd\ exhibits the smallest rate of period change among these objects. At the same time, however, the uncertainty with which $\dot{P}$ is known ($\delta\dot{P}= 1.9\times10^{-11}= 8\times10^{-7}$\,s cycle$^{-1}$) is very small. In this respect, the case of \hd\ is comparable to the cases of the best-observed magnetic CP stars exhibiting period variations,  $\sigma$\,Ori\,E and CU\,Vir \citep[see][]{Pyp20, Mik16}. The small uncertainty is due to the very short rotation period of $P=0\fd475\,5$, the time span of the observations, the relatively large amplitude of the light curve that exhibits a well defined narrow minimum (see Fig.\,\ref{krivky}), and, finally, the quality of the employed photometry. 
Such small value of $\dot{P}$ would be hardly detectable in the vast majority of known CP2 stars, which raises the question of whether the periods of all magnetic CP stars are actually variable to some extent, and on timescales shorter than justifiable by stellar evolution (the moment of inertia varies with evolution on the main sequence, resulting in a change of the rotational period by mere conservation of angular momentum).

\subsubsection{Can the period variations be explained in purely kinematic terms?}
It is not certain {\it a priori} that the secular period change is intrinsic to the star. It might possibly be due to orbital motion in a long-period binary system, to acceleration linked with the motion within the Vela OB2 association, or even to a radial acceleration due to the change of orientation of the line-of-sight linked with the proper motion. In this section, these possibilities are shortly discussed. 

First, there is a link between $\dot{P}$ and $\dot{V_\mathrm{r}}$, where $V_\mathrm{r}$ is the radial velocity, through the classical
Doppler effect formula:

\begin{align}
&P\equiv \frac{1}{\nu} \implies \frac{\mathrm{d}P}{\mathrm{d}t}=-\frac{1}{\nu^2}\frac{\mathrm{d}\nu}{\mathrm{d}t},\\
&\frac{V_\mathrm{r}}{c}=\frac{\Delta\nu}{\nu} \implies 
\frac{1}{c}\frac{\mathrm{d}V_\mathrm{r}}{\mathrm{d}t}
=\frac{1}{\nu}\frac{\mathrm{d}(\Delta\nu)}{\mathrm{d}t}-\frac{\Delta\nu}{\nu^2}\frac{\mathrm{d}\nu}{\mathrm{d}t}.
\end{align}

The right-hand side of the last equation disappears because the intrinsic frequency of the star is stable by hypothesis, so we are left with:
\begin{equation}
 \frac{\mathrm{d}V_\mathrm{r}}{\mathrm{d}t}=\frac{c}{\nu}\frac{\mathrm{d}(\Delta\nu)}{\mathrm{d}t}=-c\nu\frac{\mathrm{d}P}{\mathrm{d}t}= -c\frac{\dot{P}}{P}
\end{equation}
Inserting the relevant numbers, we obtain
\begin{displaymath}
    \dot{V_\mathrm{r}}\approx 5.4\,\kms\ \mathrm{per~century} = 2.2\,\kms\  \mathrm{over~40~years.}
\end{displaymath}
With a proper motion of $9.38$~mas\,yr$^{-1}$, the tangential velocity of \hd\ is similar to 
its radial velocity, and the change of orientation of the line-of-sight is so small (assuming a constant spatial velocity vector relative to the Sun) that the change of radial velocity would amount to no more than a few cm\,s$^{-1}$.

Second, the motion of the star within the Vela OB2 association is too slow to provide any significant acceleration ($\sigma_\mathrm{V}\approx 4.5$~\kms; \citealt{2020MNRAS.493.2339M}), since the crossing time exceeds tens of thousands of years (for the Trapezium in Orion, which represents a minimal value, one finds $t_\mathrm{cross}\approx 13$\,kyr; \citealt{2006MNRAS.373..295P}), a value that much exceeds the time base of $40$~yr.

Third, assuming that \hd\ has a $1$~M$_\odot$ companion in a binary system with a circular orbit and an orbital period of $200$~yr, its orbital velocity would amount to about $2$\,\kms. For an orbital inclination close to $90^\circ$ and an orbital phase where the radial velocity is close to the systemic one (and hence varies almost linearly), $V_\mathrm{r}$ will change by about $2.3$\,\kms\ during $40$ years\footnote{We note that increasing the orbital period to $400$~yr would lead to a change in $V_\mathrm{r}$ of $\sim 1$~\kms\ at most over the same time span of $40$~yr. On the other hand, a shorter orbital period could not accommodate the time span of $40$~yr during which the radial velocity varies quasi linearly.}. This is very close to the radial acceleration needed to explain $\dot{P}$. On the other hand, the semi-major axis of the primary orbit would amount to $13.6$~au, corresponding to $33$~mas on the sky. The $Gaia$ DR3 data cover $34$~months, i.e. $1.42$\% of the orbit, which would correspond to a motion of $\sim 3$~mas on the sky -- a value easily measurable by $Gaia$, although this might be interpreted as a mere contribution to the proper motion. However, the "non-single star" parameter in the DR3 data is null for \hd, which casts doubt on the idea that $\dot{P}$ has been caused by orbital motion. Moreover, the "Renormalised Unit Weight Error" parameter $RUWE=1.08$ indicates that the standard single-star model fits the $Gaia$ data well (doubt arises only in the case of $RUWE > 1.4$).

\subsubsection{Possible intrinsic cause}
An interpretation of the period change observed in magnetic CP stars is far from obvious. In cases of period lengthening, one may expect that the latter is due to loss of angular momentum linked with mass loss where the stellar wind flows along open magnetic lines. Such a scenario has been proposed for $\sigma$ Ori E \citep{2009MNRAS.392.1022U, 2010ApJ...714L.318T}. The observed characteristic spin-down time of this $9$~M$_\odot$ star is $\tau_\mathrm{spin}=1.34$~Myr \citep{2010ApJ...714L.318T}, which is consistent with theoretical predictions \citep{2009MNRAS.392.1022U} given a mass loss rate of $\dot{M}=2.4\times 10^{-9}$~M$_\odot$\,yr$^{-1}$ estimated through modelling \citep{2006A&A...460..145K}. 

This star, however, is three times more massive than \hd\ with a luminosity of $\log(L/L_\odot)=3.5$ \citep{2019MNRAS.485.1508S}, while the luminosity of \hd\ is $\log(L/L_\odot)=2.0$ (adopting $R$ and $T_\mathrm{eff}$ from Table~\ref{atmospheric_parameters}). A rough extrapolation using Fig.~1 of \citet{1988A&AS...72..259D} indicates that such a low luminosity suggests a mass loss of $\sim 10^{-12}$~M$_\odot$\,yr$^{-1}$ at most, and this is much higher than the $\sim 10^{-16}$ value proposed by \citet{Babel95} on a theoretical basis. Assuming that Equation~25 of \citet{2009MNRAS.392.1022U} is valid for the low effective temperature of \hd\ (which is admittedly debatable) and inserting the very high value of $10^{-12}$~M$_\odot$\,yr$^{-1}$, we get $\tau_\mathrm{spin}\sim 520$~Myr (adopting $M=3$~M$_\odot$, $R=2$~R$_\odot$, $k=0.1$, $B_\mathrm{p}=1$~kG, and $v_8\equiv v_{\infty}/1000\,\mathrm{km\,s}^{-1}=1$). This is almost two orders of magnitude longer than the observed value of  $\tau_\mathrm{spin}=5.5\pm 0.4$~Myr. In summary, it seems difficult to explain the observed period lengthening through stellar wind.

Moreover, for a significant fraction of period-changing magnetic CP stars, an acceleration or cyclic changes of the rotation period are observed \citep[see in][]{Mik21}. This can be interpreted within the model of torsional oscillations \citep{2017MNRAS.464..933K,takahashi21}. This model, however, requires an internal magnetic field with strength equal to the surface field to explain the observed time scales of period variability, which contradicts the current understanding of the internal magnetic field of magnetic CP stars. It is generally admitted that the magnetic field of these objects is a fossil one \citep{2004Natur.431..819B, 2019Natur.574..211S}, and that the strength of such a field increases with depth, as observed by e.g. \citet{2022MNRAS.512L..16L}. It is not excluded that torsional oscillations may affect only outer layers frozen in the outer magnetic field \citep{mik18}, but this idea has not been theoretically studied up to now.

\subsection{Rotational flattening, equatorial velocity, and inclination of the rotational axis}\label{parameters}

The simplest approximation for the shape of a uniformly rotating single star is provided by the Roche model, which accounts for the centripetal gravitational force and the centrifugal force of rotation. Following this approach, the stellar surface has the shape of an isobar surface expressed by the sum of the gravitational potential approximated by the field of a point mass with mass $M$ of the star and the centrifugal potential of a rigid body rotating with angular velocity \Om. This isobar area is an axisymmetric spheroid characterised by the equatorial and polar radii \Req, \Rp\ and flattening $f$, $f=\Req/\Rp$. After some algebra, we obtain the basic relation between the radii from Eqs. (10) and (12) of \citet{Zahn10}, as follows:
\begin{align}
f=&\,\frac{\Req}{\Rp}=1+\frac{\Om^2\Req^3}{2\,GM}=1+\frac{\Om^2\Rp^3}{2\,GM}\,f^3= 1+\frac{\Om^2}{2\,\Om_{\mathrm{crit}}^2}=\label{Roche}\\
=&\,1+\frac{V^2_{\mathrm{eq}}}{2\,V_{\mathrm{crit}}^2}=1+\frac{1}{2}\zav{\frac{a_{\mathrm{cfg}}}{g}}_{\mathrm{eq}},\nonumber
\end{align}
where $G$ is the gravitational constant and $\Om_{\mathrm{crit}}$ the maximum angular velocity $(\Om\leq\Om_{\mathrm{crit}})$, beyond which the stellar body is disrupted. $V_{\mathrm{eq}}$ is the equatorial velocity, $V_{\mathrm{eq}}=\Req\,\Om$, while $V_{\mathrm{crit}}$ is the maximum allowed equatorial velocity ($V_{\mathrm{eq}}\leq V_{\mathrm{crit}}$). $(a_{\mathrm{cfg}}/g)_{\mathrm{eq}}$ is the ratio of the centrifugal and gravitational accelerations at the equator. It follows from Eq. (\ref{Roche}) that the maximum flattening $f_{\rm{crit}}$ of a rotating body is $f_{\rm{crit}}=\textstyle{\frac{3}{2}}$, because $a_{\mathrm{cfg}}=g$ in this case.  Critical quantities for the given mass $M$ and polar radius \Rp\ are as follows (with $\rho\equiv \Req^{\mathrm{crit}}$):
\begin{align}
    &\rho=\frac{3}{2}\,\Rp;\quad\Om_{\mathrm{crit}}=\sqrt{\,\frac{G M}{\rho^3}};\quad P_{\mathrm{min}}=\sqrt{\,\frac{4\,\pi^2\rho^3}{G M}};\label{critical}\\
    &V_{\mathrm{crit}}=\sqrt{\,\frac{G M}{\rho}}=\rho\,\Om_{\mathrm{crit}};\quad
     a_{\mathrm{cfg\,crit}}=g_{\mathrm{eq\,crit}}=\frac{G M}{\rho^2}.\nonumber
\end{align}

Applying relations (\ref{Roche}) and (\ref{critical}) to the case of \hd, we assume a mass of $M=3.1\,\Ms$,  $\mathit{\Omega}=1.5294\times10^{-4}$\,s$^{-1}$, and $\Rp\doteq R=1.97\,\Rs$. After several iterations of these semianalytic relations between $R_{\rm{eq}}$ and $R_{\rm{p}}$
\citep{ekstrom08}, we estimate the equatorial radius at $\Req=2.16\,\Rs$. Bearing in mind that the rotational period of \hd\ is the smallest among all known CP2 stars, the derived flattening of $f = 1.096$ appears surprisingly small at first glance. That is because the flattening $f$ and the ratio of the equatorial centrifugal to gravitational accelerations ($(a_{\mathrm{cfg}}/g)_{\mathrm{eq}}=0.19$ for \hd) of a rotating star critically depend on its radius ($\sim R^3$). As our target star is nearly on the ZAMS (cf. Section \ref{physta}), its polar radius is the smallest possible for its mass. 

The same conclusion follows from a comparison of the real equatorial velocity  $V_{\mathrm{eq}}=50.6\,\Req/P_1=230\,\kms$\ and the maximum allowed equatorial velocity $V_\mathrm{crit}=450\,\kms$.\footnote{It is worth noting that the much more sophisticated models of rotating stars by \citet{2013A&A...553A..24G}, which account for details of the inner structure, yield almost identical estimates of the astrophysical parameters of \hd, resulting in $V_\mathrm{crit}= 460\,\kms$.} The true rotational velocity $V_{\mathrm{eq}}$ is about half the critical one. 

We can also evaluate the quantity $\Req \sin i=\vsini\  P_1/50.6=1.79\pm0.19\,\Rs$, determining the lower limit of the stellar equatorial radius \Req. We use the simplistic approximation $R_{\rm{eq}}\approx \langle R\rangle$ according to the relatively small ratio $f=1.096$. Assuming that $\Req=2.16\,\Rs$, we derive $\sin i=0.83(9)$ which indicates an inclination angle of $i=56^{\circ}\pm9^{\circ}$.

\subsection{Interpretation of the light variability}\label{LC_inter}

As shown by \citet{Krt07}, the main cause of rotationally modulated light changes in chemically peculiar stars is their uneven distribution of overabundant chemical elements across the stellar surface. In the temperature range of \hd, the most important element in this respect is Si \citep[][]{Krt09,Krt12}, which has a number of absorption lines and bound-free transitions in the far ultraviolet (UV) region that cause a massive redistribution of far UV flux to near UV and optical regions. This mechanism, also known as blanketing, affects the entire mentioned spectral region; however, the intensity of the flux redistribution decreases with increasing wavelength \citep{Krt15}. This is totally in line with the observed light variability pattern of \hd, which is charaterized by decreasing amplitude towards longer wavelengths (see Figs.\,\ref{amplitudy} and \ref{krivky}).

Following this scenario, the areas of increased Si abundance appear brighter at optical and infrared wavelengths than those with a lower content of this element. Therefore, during the time of maximum brightness, bright Si spots are expected to cover the major part of the visible stellar hemisphere. In this respect, it is noteworthy that the spectropolarimetric measurement used for the abundance analysis covers the rotational phases 0.015\:--\:0.040 and was therefore taken almost directly during minimum light. Consequently, the derived overabundance of Si (and perhaps that of Fe, if the surface distribution of both elements agree) may in fact be significantly smaller than the average overabundance.

\subsection{Development of chemical peculiarity during the main-sequence stage}

With a derived age of 10 Myr (Sect. \ref{physta}), \hd\ is a very young object that has only spent about 1/27 of its expected main-sequence life time \citep[365 Myr for 3.1\,\Ms, according to][]{Ekstr12}. Despite this short time, radiative diffusion has obviously efficiently established a peculiar atmospheric composition, in agreement with findings for other young stars \citep{Rom20,Rom15}.

The subsequent rotational evolution of \hd\ depends on how angular momentum is transported within the star \citep{Kesz20}. The non-magnetic evolutionary models of \citet{granada}, which use the horizontal diffusion coefficient of \citet{zahn92} and the shear diffusion coefficient of \citet{maeder97}, predict a nearly constant equatorial rotational velocity on the main sequence. Following this scenario, \hd\ should maintain its rotational speed and possibly also its chemical peculiarity during main-sequence evolution.

On the other hand, for magnetic stars, it may be more reasonable to assume solid-body rotation \citep{takala}. Neglecting angular momentum loss due to the magnetised wind, which is a reasonable assumption for a star with a temperature of \hd\ \citep{metuje}, the rotational period is predicted to remain nearly constant during the main-sequence lifetime \citep{takala}, thus  $\Om(t)=\Om_0=$\,const. As the equatorial radius $\Req(t)$ increases during the main-sequence evolution, the stellar equatorial velocity $V_{\rm{eq}}(t)=\Om_0\,\Req(t)$ may approach the critical rotation limit $V_{\rm{crit}}$ (see Eq. \ref{critical}), which results in the loss of the peculiar atmospheric abundance pattern.

The subsequent spinning up of stars during the main-sequence evolution could be at the root of why so few ultra short-period magnetic CP stars are observed.

\section{Conclusions}

Using high-resolution spectroscopy and an extensive set of ground- and space-based time series photometry, we carried out a detailed study of the CP2 star \hd. Our main results are summarised as follows.

\begin{enumerate}
   
   \item With an age of only 10 Myr, an effective temperature of $\Teff=13\,000\pm300$\,K, surface gravity log\,$g$\,=\,$4.10\pm0.10$, radius $R=1.97\pm0.09\,\Rs$, and mass $M=3.1\pm0.1\,\Ms$, \hd\ is situated close to the zero-age main sequence and a member of the open cluster NGC 2547 in the Vela OB2 complex.
   
   \item We estimated the flattening of the star at a value of $1.096$, which means that the star is nearly spherical. This agrees with the fact that the equatorial velocity $V_{\mathrm{eq}}=50.6\,R/P=230$\,\kms\ is half the calculated critical equatorial velocity $v_\mathrm{crit}= 460$\,\kms. The observed quantities of $R\sin i=1.79\pm0.19$\,R$_\odot$ and $\sin i=0.83\pm0.09$ imply an inclination angle of $i=56^{\circ}$.
   
   \item \hd\ is a classical late B-type CP2 star showing strong overabundances of Mg (1.8~dex), Si (1.9~dex), Ca (1.6~dex), Ti (2.2~dex), and Fe (1.8~dex). We showed that within the errors, a dipolar magnetic field of the order of 1\,kG does not significantly affect the abundance determination.
   
   \item In agreement with this classification, we find spectroscopic evidence for the presence of chemical surface spots, which are responsible for the observed light changes. Si, in particular, is expected to be a major contributor to the photometric variability.
   
   \item No conclusive evidence for the presence of a strong magnetic field was found in the available spectroscopic data. A typical Zeeman signature was found in one ESPaDOnS spectrum, from which we derive a longitudinal magnetic field of $\langle B_z \rangle = 270\pm150$\,G. However, from a statistical point of view, this signal is not significant, which implies that the magnetic field is either very weak or could not be detected because of insufficient data quality or unfortunate phase coverage. 
   
   \item We confirm \hd\ as the CP2 star with the shortest known rotational period ($P=0\fd4755$). We found the available multi-color light curves from 1981-2021 covering the wavelength range 3\,400\:--\:8\,000\,\AA\ to be constant in shape and amplitude, which made it possible to apply the phenomenological model used, Eq. (\ref{model}) and Appendix \ref{LC model}.
   
   \item After careful treatment of the photometric data, we conclude that the rotational period of \hd\ is slowly lengthening $(\dot{P}= 2.36(19)\times10^{-10}=7.4(6)\,\textrm{ms yr}^{-1})$. The best fit to the available data is provided by the following quadratic ephemeris: HJD$_\mathrm{min}(E)=2\,459\,212.969\,35+0\fd475\,516\,64\,E+5\fd62\times10^{-11}E^2$.
   
   \item Although the derived value of $\dot{P}$ is the smallest rate of period change known among magnetic chemically peculiar stars, the detection is highly significant. It was made possible thanks to the shortness of the period, the long time span and high quality of the time series photometry, and a light curve with relatively large amplitude and a well-defined narrow minimum.

\end{enumerate}
In summary, \hd\ is an outstanding object that may ultimately turn out to be a keystone in the understanding of CP2 star evolution. Theory needs to explain the establishment and maintenance of chemical peculiarities in such a young and fast rotating object.

The phenomenon of period variability may be inherent to all magnetic chemically peculiar stars. However, only further detailed (and hence time- and resource-consuming) studies will be able to shed more light on this subject. As yet, we can offer no consistent explanation of this phenomenon.

\begin{acknowledgements}
This work has been supported by the Erasmus+ programme of the 
European Union under grant number 2020-1-CZ01-KA203-078200.
PLN thanks Dr. Laurent Eyer for his help in recovering the Geneva
photometric measurements of HD 60431. This work is based on observations obtained at the Canada-France-Hawaii Telescope (CFHT) which is operated by the National Research Council of Canada, the Institut National des Sciences de l'Univers of the Centre National de la Recherche Scientique of France, and the University of Hawaii and on observations made with MRES at the Thai National Observatory, which is operated by the National Astronomical Research Institute of Thailand (Public Organization). 
\end{acknowledgements}

\bibliographystyle{aa}

\bibliography{aa}

\appendix
\section{Light curve model}\label{LC model}
The normalised function $B(\vartheta,\vec{\alpha})$ presented in Eq.~(\ref {model}), which expresses the appearance of phased light curves in all observed passbands, is defined as a linear combination of nine mutually orthonormal functions with alpha parameters. Five of them are simple symmetric functions with a minimum at phase 0 and four are antisymmetric functions with zero derivative at phase 0. 
\begin{align}
   & B(\vartheta,\vec{\alpha})=-\sum_{i=1}^5 \alpha_i \cos(2\pi i\vartheta)-\sum_{i=1}^4 \alpha_{i+5}\sum_{j=1}^{i+1}\beta_{ij}\sin(2\pi j\vartheta);\label{LCmodel}\\
   & \sum_{j=1}^{i+1} j\,\beta_{ij}=0;\quad \sum_{j=1}^{i+1}\beta_{ij}\  \beta_{kj}=\delta_{ik},\ \mathrm{for}\ i\geq k; \label{constr.}
 \end{align}
Parameters $\beta_{ij}$ that fulfill the constraints (\ref{constr.}) are the following: $\beta_{11}=0.89443$, $\beta_{12}=-0.44721$; $\beta_{21}=0.35857$, $\beta_{22}=0.71714$, $\beta_{23}=-0.59761$; $\beta_{31}=0.19518$, $\beta_{32}=0.39036$, $\beta_{33}=0.58554$, $\beta_{34}=-0.68313$; $\beta_{41}=0.12309$, $\beta_{42}=0.24618$, $\beta_{43}=0.36927$, $\beta_{44}=0.49237$, and $\beta_{45}=-0.7185494$. 

The coefficients of the function  $B(\vartheta,\vec{\alpha})$ must satisfy the normalisation condition, 
\begin{equation}
    \sum_{i=1}^{9} \alpha_{i}^2=1;\quad \Rightarrow\quad \alpha_1=\sqrt{1-\sum_{i=2}^{9}\alpha_i^2} \label{norm}\\
\end{equation}
thus we are only looking at eight components of the $\vec{\alpha}$ vector to describe $B(\vartheta,\vec{\alpha})$.

\begin{table}
\caption{Found coefficients $\alpha_i$ describing the appearance of \hd\ light curves (see Eq. (\ref{LCmodel}) and (\ref{norm})}
\centering
\begin{tabular}{ll}
\hline
Symmetric terms & Antisymmetric terms\\
\hline
$\alpha_1=0.9543$ & \\
$\alpha_2=0.2584(7)$ & $\alpha_6=0.1420(14)$\\
$\alpha_3=0.0205(5)$ & $\alpha_7=-0.0223(18)$\\
$\alpha_4=-0.0089(6)$ & $\alpha_8=-0.0359(10)$\\
$\alpha_5=0.0022(5)$ & $\alpha_9=-0.0044(7)$\\
\hline
\end{tabular}\label{alpha}
\end{table}
\begin{figure}
\begin{center}
\includegraphics[width=0.40\textwidth]{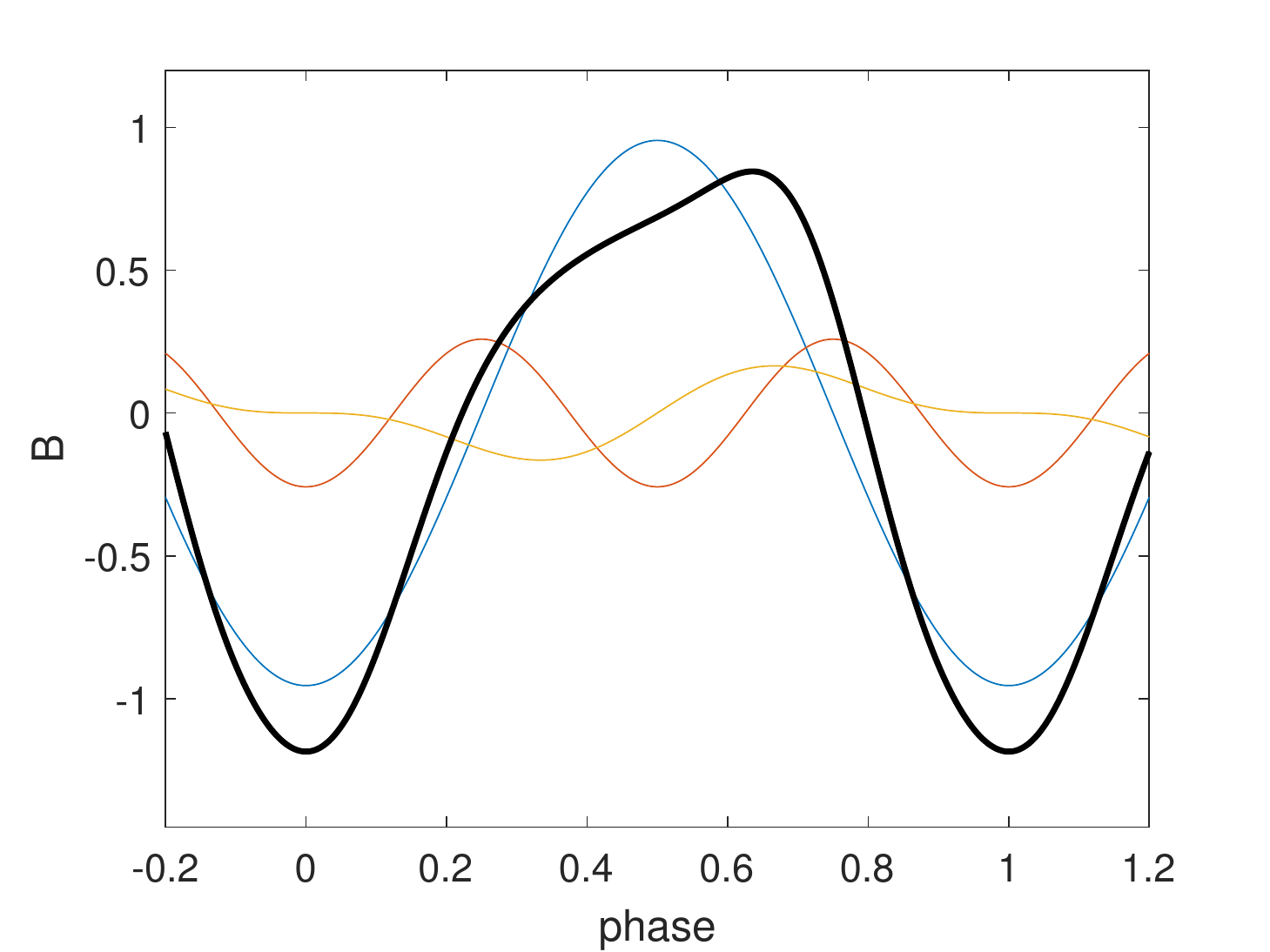}
\caption{Resulting normalized function $B(\vartheta,\vec{\alpha})$ (bold line) and  three  most intensive components of it.}
\label{slozky}
\end{center}
\end{figure}

{\section{Normal points}\label{normalka}

For a detailed representation of complex light variations defined by a set of individual photometric observations $\{t_i,y_i,w_i\}$, where $t_i$ is the moment or phase of the observation, $y_i$ is the measured quantity (magnitude or intensity), and $w_i$ is the weight of the $i$-th observation, it is helpful to indicate the course of the mean light curve using so-called ‘normal points‘, which can be calculated as follows.

First, the set of observations is roughly fit by a suitable model function of time or phase $f(t)$. To this end, we usually use polynomials or harmonic polynomials of low orders. In the case of outlier-adjusted data, standard regression can be applied; otherwise, it is desirable to use robust regression that eliminates the effects of outliers. For each observation, we calculate the residual of the measured quantity to the selected model $\Delta y_i = y_i-f(t_i)$.

{Next, we sort the set data, supplemented by $\Delta y_i$ residue values, according to time or phase and divide it into $N$ subsets with sufficient observations. For each subset, we determine the weighted arithmetic mean of the residuals $\Delta Y_j$ and the mean weighted squared deviation of this mean $\delta Y_j $, which is also the uncertainty of determining the value of the $j$-th normal point.

In the next step, for each of the subsets, we determine the centre $\tau_j$ as the weighted arithmetic average of the observation times $t_{ij}$ in the subset. Normal point value $Y_j = f(\tau_j) + \Delta Y_j$. After that, the normal points can be plotted in a graph, preferably against the background of a set of individual observations. We change the number of normal points $N$ and modify the interleaved supporting mathematical model $f(t)$ depending on how the set of normal points follows the perceived mean light curve.}}

\end{document}